\documentclass{article}
\usepackage{graphicx}  
\usepackage{amsmath}   
\usepackage[compress]{cite}
\usepackage{amssymb}   
\usepackage{bm} 
\usepackage{dcolumn}
\usepackage{color}
\usepackage{mathrsfs}
\usepackage{amsfonts}
\RequirePackage[colorlinks,citecolor=blue,urlcolor=magenta,linkcolor=blue]{hyperref}
\def\eq#1{{Eq.~(\ref{#1})}}
\def\eqs#1{{Eqs.~(\ref{#1})}}
\def\sect#1{{Sec.~\ref{#1}}}
\def\EH{Einstein-Hilbert }

\title{Effective gravitational field equations on m-brane embedded in n-dimensional bulk of Einstein and f(R) gravity}
\author{Sumanta Chakraborty 
\footnote{sumanta@iucaa.in;~~sumantac.physics@gmail.com}
and 
Soumitra SenGupta
\footnote{tpssg@iacs.res.in}\\
{\small {IUCAA, Post Bag 4, Ganeshkhind}}\\
{\small {Pune University Campus, Pune 411 007, India}}\\
and\\
{\small{Department of Theoretical Physics}}\\
{\small {Indian Association for the Cultivation of Science, Kolkata-700032, India}}}
\begin{document}
  
\maketitle

\begin{abstract}
We have derived effective gravitational field equations on a lower dimensional hypersurface (known as a brane), placed in a higher dimensional bulk spacetime for both Einstein and $f(\mathcal{R})$ gravity theories. We have started our analysis on $n$-dimensional bulk from which the effective field equations on a $(n-1)$-dimensional brane has been obtained by imposing $Z_{2}$ symmetry. Subsequently, we have arrived at the effective equations in $(n-2)$-dimensions starting from the effective equations for $(n-1)$ dimensional brane. This analysis has been carried forward and is used to obtain the effective field equations in $(n-m)$-dimensional brane, embedded in a $n$-dimensional bulk. Having obtained the effective field equations in Einstein gravity, we have subsequently generalized the effective field equation in $(n-m)$-dimensional brane which is embedded the $n$-dimensional bulk spacetime endowed with $f(\mathcal{R})$ gravity. We have also presented applications of our results in the context of Einstein 
and $f(\mathcal{R})$ gravity. In both the cases we have discussed vacuum static spherically symmetric solutions as well as solutions in cosmological context. Implications are also discussed.
\end{abstract}
\section{Introduction}\label{Sec:Intro}

One of the fundamental problems in theoretical physics amounts to unifying all known interactions. A consistent unification of gravity with the other fundamental interactions has been an active area of research over many years. One of the leading contenders along this direction is superstring theory. In superstring theory the universe is assumed to be 11-dimensional of which seven are compactified leaving four non-compact dimensions, which we observe. Among various candidates, the 10-dimensional $E_{8}\times E_{8}$ heterotic string theory is a strong candidate as the theory is able to accommodate the standard model within it. Recently it has been shown that the 10-dimensional $E_{8}\times E_{8}$ heterotic string theory is related to the 11-dimensional M theory (with an orbifold symmetry $S^{1}/Z_{2}$). This also confines the standard model particles to standard non-compact 4-dimensional spacetime while gravity can probe the higher dimensional spacetime \cite{Polchinski1998,Horava1996a,Horava1996b}.

A simplistic model which captures most of the key notions is a 5-dimensional model in which matter fields are confined to 4-dimensional spacetime while gravity \emph{exists} in all the five spacetime dimensions. The first step along this direction with this simplified picture presented above was taken by Randall and Sundrum by introducing two 4-dimensional sub-manifolds (known as brane) in a 5-dimensional anti-de Sitter bulk \cite{Randall1999a}. This arrangement of two positive and negative tension branes separated by a finite distance (known as radion field \cite{Csaki2000,Goldberger1999,Chakraborty2014a}) can address the hierarchy problem by lowering the Higgs mass to electroweak scale on the visible brane. Subsequently, they have introduced another model, which consists of a positive tension brane with an infinite extra dimension \cite{Randall1999b}. Due to the presence of an extra spacetime dimension it is expected that there would be deviation of various physical results from Einstein gravity, which 
would become more significant in the high energy limit. The brane world model subsequently, was applied in the context of particle phenomenology \cite{Djouadi2008,Davoudiasl2000,Davoudiasl2001,Hundi2013}, black hole physics \cite{Chamblin2000,Garriga2000,Lukas1999,Dadhich2000,Chakraborty2008} and cosmology \cite{Chakraborty2014a,Binetruy2000,Csaki1999,Cvetic1997,Benakli1999}.

A natural extension of the Randall-Sundrum (RS) scenario to models with more than one warped extra dimension have been proposed, where several independent $S^{1}/Z_{2}$ orbifold is considered along with the 4-dimensional non-compact manifold $M_{4}$ \cite{Chacko2000,Cohen1999,Gregory2000,Giovannini2001,Choudhury2007}. These multiply warped models are mainly motivated from various considerations, which include: radion stabilization, fermion mass hierarchy in Standard Model etc. There have been significant number of follow-up works in these multiply warped models which involves: inflation, cosmic acceleration, matter field localization, Kaluza-Klein modes of graviton, gauge and scalar fields \cite{Mcdonald2007,Saharian2006,Koley2008,Das2011,Mukhopadhyaya2013,Chakraborty2014b}. However all these results depend on certain assumptions of bulk metric elements and bulk Einstein equation under some simplified conditions, for example in presence of homogeneity and isotropy. 

There exist an elegant way in which effective field equation for gravity on a lower dimensional spacetime can be obtained. This involves use of geometric quantities like metric, curvature induced on the brane from bulk. Using these induced geometric quantities it is possible to  arrive at a relation between the bulk curvature tensor and the brane curvature tensor, known as Gauss-Codazzi equation \cite{Padmanabhan2010,Poisson2004}. From various contractions of the Gauss-Codazzi equation we can determine the effective field equation on the brane by relating bulk Einstein tensor to that on the brane. 

This approach was applied in order to obtain the gravitational field equation on the brane from a 5-dimensional bulk in \cite{Shiromizu2000}. The effective equation contains non-local terms inherited from the bulk. Subsequently starting from the effective field equation derived in \cite{Shiromizu2000} a class of vacuum solutions has been obtained in \cite{Dadhich2000}, which has been further generalized in \cite{Harko2004}. 

It is strongly believed that general relativity is only a low energy approximation of some underlying fundamental high energy theory \cite{Buchbinder1992,Vassilevich2003}. This suggests that Einstein-Hilbert action in high energy limit, should be modified by introduction of higher curvature terms. Such an alternative theory is the $f(\mathcal{R})$ theory, where $\mathcal{R}$ is the bulk Ricci scalar. This theory has been studied extensively in the context of solar system tests, inflationary paradigm, late time cosmic acceleration along with possibility of detecting gravitational waves \cite{Nojiri2011,Sotiriou2010,Nojiri2003,Felice2010,Corda2009}. Moreover two-brane models, due to the presence of $f(\mathcal{R})$ term in the bulk has also been a topic of discussion recently in collider physics \cite{Aslan2006,Silva2013,Chakraborty2014c} relating to the absence of graviton Kaluza-Klein Modes in LHC. 

The above approach of obtaining effective gravitational field equation is usually confined to Einstein gravity on the 3-brane alone. This subsequently has been generalized to obtain effective field equation on the 3-brane when the bulk contains $f(\mathcal{R})$ gravity. This effective field equations has been used in \cite{Chakraborty2015a} to obtain vacuum solution in $f(\mathcal{R})$ bulk. 

In this work, we generalize the above setup by working in an $n$-dimensional bulk and then obtaining effective equation on an $(n-m)$-dimensional brane where $m$ can take any values. This seems natural from a string theory view point, where the universe is 11-dimensional, while our universe is a 3-brane with 4-spacetime dimensions. Following this argument we have derived effective gravitational field equation on a $(n-m)$-dimensional brane embedded in a $n$-dimensional bulk. We start with a $n$-dimensional bulk spacetime and then derive the effective field equation on a $(n-1)$-dimensional brane, which can then be used as the starting point to derive effective field equation on a $(n-2)$-dimensional brane. Having obtained the effective field equation on the $(n-2)$-dimensional brane one can easily recognize the pattern of obtaining the effective field equation in any arbitrary lower dimensional surfaces. Thus following the road from $n$-dimensional bulk to $(n-2)$-dimensional brane, we can obtain the 
effective field equation on $(n-m)$ dimensional brane as well. The above analysis can be easily generalized to $f(\mathcal{R})$ gravity model in the bulk following the analysis presented in \cite{Chakraborty2015a}. 

The paper is organized as follows: In \sect{Sec:Effective} we have elaborated the derivation of effective field equation for Einstein gravity. First, in \sect{Sec01:n-1} we present the derivation of effective field equation in $(n-1)$-dimensional brane, which is subsequently generalized in \sect{Sec02:n-2} to $(n-2)$-dimension. Following the pattern we obtain the effective gravitational field equation in $(n-m)$-dimensional brane in \sect{Sec03:n-m}. This result has been generalized for $f(\mathcal{R})$ gravity model in \sect{Sec04:f(R)}. The effective equations so derived have been applied to obtain static spherically symmetric vacuum solutions and cosmological solutions along with their compatibility with standard cosmological evolutions. The results in the context of Einstein gravity have been discussed in \sect{AppGR} while those for $f(\mathcal{R})$ gravity is discussed in \sect{Sec:Application}. Finally we conclude with a discussion on the results obtained.

Throughout the paper, we maintain the following convention: all Latin letters $a,b,\ldots$ run over all the $n$ spacetime indices, Greek letters $\alpha ,\mu ,\ldots$ run over the $(n-1)$ spacetime indices. Finally capitalized Latin letters $A,B,\ldots$ run over the $(n-2)$ spacetime indices.
\section{Effective Field Equation In Einstein Gravity: Background and Formulation}\label{Sec:Effective}

Hamiltonian formulation in general relativity comes up with the $(1+3)$ splitting of the spacetime, where a time direction is singled out. For this purpose we introduce a time coordinate $t=t(x^{i})$ and foliate the spacetime with such $t=\textrm{constant}$ surfaces. Introducing a vector field $\mathbf{t}=\partial /\partial t$ we can move out of the surface by moving along the integral curves for $\mathbf{t}$. This involves a change in the $t$ coordinate introducing the shift function $N$ and change in the orientation captured by the lapse functions $N^{\alpha}$. Then $N,N^{\alpha}$ yield 4 components of the Einstein tensor, while the remaining six components are being determined by induced metric on the surface $h_{\alpha \beta}$. The induced metric can be obtained from the bulk metric $g_{ab}$ in two ways. First, we can introduce normalized normals to the surface given by: $n_{a}=-N\nabla _{a}t$ and then define the induced metric as $h^{ab}=g^{ab}-n^{a}n^{b}$ (This formulation does not work for null 
surfaces, for which a complete treatment has been provided in \cite{Chakraborty2015b}). Secondly, we can introduce coordinates $y^{\alpha}$ on the $t=\textrm{constant}$ hypersurface on which we can introduce surface tetrads $e^{a}_{\alpha}=\partial x^{a}/\partial y^{\alpha}$ and obtain induced metric as: $h_{\alpha \beta}=g_{ab}e^{a}_{\alpha}e^{b}_{\beta}$. With the induced metric we can define covariant derivative on the surface and then using vectors on the $t=\textrm{constant}$ surface we can introduce curvature tensor on the spatial hypersurface. The relation between 4-dimensional and 3-dimensional curvature tensor is obtained through Gauss-Codazzi equation. This can be manipulated in various ways in order to arrive at the effective Einstein's equation on the $t=\textrm{constant}$ hypersurface. This procedure can be extended easily for timelike surfaces (i.e. spacelike normals) as well. 
\subsection{Effective Field Equation on (n-1)-dimensional Brane}\label{Sec01:n-1}

Having described the general setup let us now consider the effective field equation on a brane embedded in a higher dimensional bulk, which is the main motivation of this work. The procedure sketched above can then be applied for a generic situation where a single $(n-1)$-dimensional brane is embedded in a $n$-dimensional bulk spacetime with Einstein gravity. For this set up the effective gravitational field equation on the $(n-1)$ dimensional brane can be written as (see App. \ref{App1}): 
\begin{align}\label{GC01}
^{(n-1)}G_{\alpha \beta}&=\kappa _{n}^{2}\frac{n-3}{n-2}\left[T_{ab}e^{a}_{\alpha}e^{b}_{\beta}
+h_{\alpha \beta}\left\lbrace \epsilon T_{ab}n^{a}n^{b}-\frac{1}{n-1}T \right\rbrace \right]
-\epsilon E_{\alpha \beta}
\nonumber
\\
&+\epsilon \left[ KK_{\alpha \beta}-K_{\alpha \mu}K^{\mu}_{\beta}
-\frac{1}{2}h_{\alpha \beta}\left(K^{2}-K_{\mu \nu}K^{\mu \nu} \right)\right]
\end{align}
The above expression connects bulk energy momentum tensor and the extrinsic curvature to the $(n-1)$ dimensional Einstein tensor. Given $T_{ab}$, along with surface contribution we can solve the above equation to derive $(n-1)$ dimensional solution. 

The above effective equation is applicable to both spacelike and timelike surfaces. However for our purpose, we will need only timelike surfaces, which are the branes. Thus the basic assumption to be followed in this section is that the $n$ dimensional spacetime can be projected on a $(n-1)$ dimensional brane with spacelike normal. The line element then takes the form:
\begin{align}\label{GC02}
ds^{2}=d\chi ^{2}+h_{\alpha \beta}dx^{\alpha}dx^{\beta}
\end{align} 
The brane is located at $\chi =0$ on this $n$ dimensional spacetime. Using the standard assumption that normal matter can exist only on the brane, the bulk energy momentum tensor can be written as:
\begin{align}\label{GC03}
T_{ab}=-\Lambda _{n}g_{ab}+S_{\alpha \beta}e^{\alpha}_{a}e^{\beta}_{b}\delta \left(\chi \right)
\end{align}
where, $S_{\alpha \beta}$ represents the brane energy-momentum tensor and $\Lambda _{n}$ is the bulk cosmological constant. As the surface $\chi =0$ contain surface energy momentum tensor, the extrinsic curvature is not continuous. The discontinuity is related to the surface stress energy tensor through the following relation:
\begin{align}\label{GC04}
S_{\alpha \beta}=\frac{1}{\kappa _{n}^{2}}
\left(\left[K_{\alpha \beta}\right]-h_{\alpha \beta}\left[K\right] \right)
\end{align}
where $\left[K_{\alpha \beta}\right]$ and $\left[K\right]$ denote jump in $K_{\alpha \beta}$ and $K$ across the $\chi =0$ surface. Thus we obtain the jump in the extrinsic curvature which after imposing $Z_{2}$ symmetry, has the expression
\begin{align}\label{GC05}
K_{\alpha \beta}^{+}=-K_{\alpha \beta}^{-}=
\frac{1}{2}\kappa _{n}^{2}\left(S_{\alpha \beta}-\frac{1}{n-2}h_{\alpha \beta}S\right)
\end{align}
From the extrinsic curvature we can contract both the indices to obtain a scalar. This scalar $K$ corresponding to each side of the $\chi =0$ hypersurface is represented by $K^{\pm}$. Therefore from \eq{GC05} it has the expression: $K^{+}=-K^{-}=-\left(\kappa _{n}^{2}S/2(n-2)\right)$. Using both the expressions for $K_{\alpha \beta}$ and $K$ we finally arrive at the following identities:
\begin{align}\label{GC06}
KK_{\alpha \beta}&=\frac{\kappa _{n}^{4}}{4}\left(\frac{1}{(n-2)^{2}}h_{\alpha \beta}S^{2}
-\frac{1}{(n-2)}SS_{\alpha \beta} \right)
\\
K_{\alpha \mu}K^{\mu}_{\beta}&=\frac{\kappa _{n}^{4}}{4}\left(S_{\alpha \mu}S^{\mu}_{\beta}
-\frac{2}{(n-2)}SS_{\alpha \beta} +\frac{1}{(n-2)^{2}}h_{\alpha \beta}S^{2}\right)
\\
K^{2}&=\frac{\kappa_{n}^{4}}{4}\frac{1}{(n-2)^{2}}S^{2}
\\
K_{\alpha \beta}K^{\alpha \beta}&=\frac{\kappa _{n}^{4}}{4}\left(S_{\alpha \beta}S^{\alpha \beta}
-\frac{2}{n-2}S^{2}+\frac{n-1}{(n-2)^{2}}S^{2}\right)
\end{align}
In order to proceed further we need to decompose the brane stress-energy tensor in two parts. One of them involves the brane cosmological constant while the other provides matter stress-energy tensor on the brane. Implementation of this separation leads to,
\begin{align}\label{GC07}
S_{\alpha \beta}=-\sigma h_{\alpha \beta}+\tau _{\alpha \beta}
\end{align}
where, $\sigma$ represents the brane tension. Next we substitute this decomposition of $S_{\alpha \beta}$ into the above expressions involving extrinsic curvature. This ultimately leads to,
\begin{align}\label{GC08}
KK_{\alpha \beta}-K_{\alpha \mu}K^{\mu}_{\beta}
&-\frac{1}{2}h_{\alpha \beta}\left(K^{2}-K_{\mu \nu}K^{\mu \nu} \right)
=\kappa _{n}^{4}\Big[-\frac{1}{4}\tau _{\alpha \mu}\tau ^{\mu}_{\beta}
+\frac{1}{8}h_{\alpha \beta}\tau _{\mu \nu}\tau ^{\mu \nu}
+\frac{1}{4(n-2)}\tau \tau _{\alpha \beta}
\nonumber
\\
&-\frac{1}{8(n-2)}h_{\alpha \beta}\tau ^{2} \Big]
+\kappa _{n}^{4}\left[\frac{(n-3)}{4(n-2)}\sigma \tau _{\alpha \beta}
-\frac{(n-3)}{8(n-2)}\sigma ^{2}h_{\alpha \beta} \right]
\end{align} 
\eqs{GC03} and (\ref{GC08}) now need to be substituted in \eq{GC01} in order to obtain the final form of the effective equation on the brane. After the substitution the effective Einstein equation on the brane takes the form
\begin{align}\label{GC09}
^{(n-1)}G_{\alpha \beta}&=-\frac{(n-3)}{(n-1)}\kappa _{n}^{2}\left(\Lambda _{n}
+\frac{n-1}{8(n-2)}\kappa _{n}^{2}\sigma ^{2}\right) h_{\alpha \beta}
\nonumber
\\
&+8\pi \left(\sigma \kappa _{n}^{4}\frac{n-3}{32\pi (n-2)} \right)\tau _{\alpha \beta}
+\kappa _{n}^{4}\pi _{\alpha \beta}-E_{\alpha \beta}
\end{align}
where we have introduced the second rank tensor $\Pi _{\alpha \beta}$ as: 
\begin{align}\label{GC10}
\pi _{\alpha \beta}=\left[-\frac{1}{4}\tau _{\alpha \mu}\tau ^{\mu}_{\beta}
+\frac{1}{8}h_{\alpha \beta}\tau _{\mu \nu}\tau ^{\mu \nu}
+\frac{1}{4(n-2)}\tau \tau _{\alpha \beta}
-\frac{1}{8(n-2)}h_{\alpha \beta}\tau ^{2} \right]
\end{align}
After rewriting \eq{GC09} can also be presented in a more compact form, which can be expressed as:
\begin{align}\label{GC11}
^{(n-1)}G_{\alpha \beta}=\kappa _{n-1}^{2}T_{\alpha \beta}
\end{align}
where the energy momentum tensor $T_{\alpha \beta}$ has the expression:
\begin{align}\label{GC12}
T_{\alpha \beta}&=-\Lambda _{n-1} h_{\alpha \beta}+\tau _{\alpha \beta}
+\frac{\kappa _{n}^{4}}{\kappa _{n-1}^{2}}\pi _{\alpha \beta}-\frac{1}{\kappa _{n-1}^{2}}E_{\alpha \beta}
\end{align}
In the above expression $\Lambda _{n-1}$ represents the $(n-1)$-dimensional cosmological constant which has the expression,
\begin{align}\label{GC13}
\Lambda _{n-1}=\frac{(n-3)}{(n-1)}\frac{\kappa _{n}^{2}}{\kappa _{n-1}^{2}}\left(\Lambda _{n}
+\frac{n-1}{8(n-2)}\kappa _{n}^{2}\sigma ^{2}\right)
\end{align}
This generalizes the expression for the effective cosmological constant on a lower dimensional hypersurface embedded in a higher dimensional bulk. For $n=5$ we retrieve fine balancing relation of the Randall Sundrum single brane model \cite{Shiromizu2000}.

Thus this formalism, with a negative $n$-dimensional cosmological constant can provide a solution to the cosmological constant problem by tuning the brane tension such that $\Lambda _{n-1}=0$. In which case the brane tension can be obtained as: 
\begin{align}
\sigma ^{2}=\frac{8(n-2)\vert \Lambda _{n}\vert}{(n-1)\kappa _{n}^{2}} 
\end{align}
Note that once the $(n-1)$-dimensional cosmological constant has been set to zero it will remain zero in all the lower dimensional branes. Also $G_{n-1}$ is the $(n-1)$ dimensional gravitational constant having the expression,
\begin{align}\label{GC14}
\kappa _{n-1}^{2}=8\pi G_{n-1}= \sigma \kappa _{n}^{4}\frac{(n-3)}{4 (n-2)} 
\end{align}
Having discussed the effective gravitational field equation in $(n-1)$-dimension, we will now move one more step by deriving effective field equation in $(n-2)$-dimensional spacetime. 
\subsection{Effective Field Equation on (n-2)-dimensional Brane}\label{Sec02:n-2}

Similar analysis can be performed while obtaining effective field equation on the $(n-2)$ dimensional brane starting from the $(n-1)$-dimensional spacetime. The field equation in the $(n-1)$ dimensional spacetime can be used which ultimately leads to the following expression for effective field equation in $(n-2)$ dimensions as (see App. \ref{App2}):
\begin{align}
^{(n-2)}G_{AB}&=\frac{(n-4)}{(n-3)}\kappa _{n-1}^{2}
\left[\mathcal{T}_{\alpha \beta}e^{\alpha}_{A}e^{\beta}_{B}
+\left\lbrace \mathcal{T}_{\alpha \beta}s^{\alpha}s^{\beta}
-\frac{1}{(n-2)}\mathcal{T} \right\rbrace q_{AB}\right]
\nonumber
\\
&-\mathcal{E}_{AB}+\left[\mathcal{K}_{AB}\mathcal{K}-\mathcal{K}_{AC}\mathcal{K}^{C}_{B}
-\frac{1}{2}q_{AB}\left(\mathcal{K}^{2}-\mathcal{K}_{CD}\mathcal{K}^{CD}\right)\right]
\end{align}
This expression represents the effective Einstein's equation in an $(n-2)$ dimensional spacetime. Our next task is to rewrite the above equation in terms of various components of energy momentum tensor, especially the extrinsic curvature components. 

Let us now try to match our solutions to that of $(n-2)$ dimensional brane energy momentum tensor. As usual we assume the following form of metric ansatz, 
\begin{align}
ds^{2}&=d\chi ^{2}+h_{\alpha \beta}dx^{\alpha}dx^{\beta}
\nonumber
\\
&=d\chi ^{2}+h_{\alpha \beta}e^{\alpha}_{\zeta}e^{\beta}_{\zeta}d\zeta ^{2}+q_{AB}dx^{A}dx^{B}
\end{align}
where $q_{AB}$ is the induced metric on the $(n-2)$-dimensional brane. Following the previous derivation of effective field equation here also we divide the energy momentum tensor, $\tau _{\alpha \beta}$ in two parts, $t_{\alpha \beta}$ pertaining to $(n-1)$ dimensional brane and $\xi _{AB}$ as energy momentum tensor in the $(n-2)$ dimensional brane. Thus we have the following relation,
\begin{align}
\tau _{\alpha \beta}&=t_{\alpha \beta}+\delta \left(\zeta \right)\xi _{AB}e^{A}_{\alpha}e^{B}_{\beta}
\nonumber
\\
&=t_{\alpha \beta}+\delta \left(\zeta \right)e^{A}_{\alpha}e^{B}_{\beta}
\left(-\Sigma q_{AB}+\nu _{AB}\right)
\end{align} 
where $\Sigma$ is the brane tension and $\nu _{AB}$ is the brane energy momentum tensor. Then the tensor $\xi _{AB}$ can be obtained from the discontinuity of the extrinsic curvature from the relation,
\begin{align}
\xi _{AB}=\frac{1}{\kappa _{n-1}^{2}}
\left(\left[\mathcal{K}_{AB}\right]-q_{AB}\left[\mathcal{K}\right] \right)
\end{align}
The jump in the extrinsic curvature by imposing $Z_{2}$ symmetry can be obtained in terms of $\xi _{AB}$ as,
\begin{align}
\mathcal{K}_{AB}^{+}=-\mathcal{K}_{AB}^{-}=
\frac{1}{2}\kappa _{n-1}^{2}\left(\xi _{AB}-\frac{1}{n-3}q_{AB}\xi \right)
\end{align}
Having obtained the extrinsic curvature, the scalar $\mathcal{K}$ can be obtained by contraction with $q_{AB}$. This has the expression: $\mathcal{K}^{+}=-\mathcal{K}^{-}=-\left(\kappa _{n-1}^{2}\xi /2(n-3)\right)$. We therefore arrive at the following expressions for various combinations of $\xi _{AB}$ as they appear in effective field equation:
\begin{align}
\xi \xi _{AB}&=(n-2)\Sigma ^{2}q_{AB}-\nu \Sigma q_{AB}-(n-2)\Sigma \nu _{AB}+\nu \nu_{AB}
\\
\xi _{AC}\xi ^{C}_{B}&=\Sigma ^{2}q_{AB}-2\Sigma \nu _{AB}+\nu _{AC}\nu ^{C}_{B}
\\
\xi ^{2}&=(n-2)^{2}\Sigma ^{2}-2(n-2)\Sigma \nu +\nu ^{2}
\\
\xi _{AB}\xi ^{AB}&=(n-2)\Sigma ^{2}-2\Sigma \nu +\nu _{AB}\nu ^{AB}
\end{align}
This enables us to write an expression for the extrinsic curvature contractions in terms of $\xi _{AB}$, which can further be simplified in terms of $\nu _{AB}$ following the above identities as:
\begin{align}
\mathcal{K}_{AB}\mathcal{K}&-\mathcal{K}_{AC}\mathcal{K}^{C}_{B}
-\frac{1}{2}q_{AB}\left(\mathcal{K}^{2}-\mathcal{K}_{CD}\mathcal{K}^{CD}\right)
\nonumber
\\
&=\frac{\kappa _{n-1}^{4}}{4}\left[\frac{1}{n-3}\xi \xi _{AB}-\xi _{AC}\xi ^{C}_{B}
-\frac{1}{2(n-3)}\xi ^{2}q_{AB}+\frac{1}{2}q_{AB}\xi ^{CD}\xi _{CD} \right]
\nonumber
\\
&=\kappa _{n-1}^{4}\left[-\frac{1}{4}\nu _{AC}\nu ^{C}_{B}+\frac{1}{8}q_{AB}\nu _{CD}\nu ^{CD}
+\frac{1}{4(n-3)}\nu \nu _{AB}-\frac{1}{8(n-3)}q_{AB}\nu ^{2} \right]
\nonumber
\\
&-\kappa _{n-1}^{4}\Sigma ^{2}\frac{(n-4)}{8(n-3)}q_{AB}
+\kappa _{n-1}^{4}\Sigma \frac{(n-4)}{4(n-3)}q_{AB}
\end{align}
It now just needs some rearrangement to write down the effective field equation on the $(n-2)$ dimensional brane. After inserting all relevant expressions and then rearranging the effective Einstein's equation on this $(n-2)$ dimensional brane we obtain,
\begin{align}
^{(n-2)}G_{AB}&=-\left(\frac{n-4}{n-2}\right)\Lambda _{n-1}q_{AB}
\nonumber
\\
&+\left(\frac{n-4}{n-3}\right)8\pi G_{n-1}\left[t_{\alpha \beta}e^{\alpha}_{A}e^{\beta}_{B}
+q_{AB}\left\lbrace t_{\alpha \beta}s^{\alpha}s^{\beta}-\frac{1}{n-2}t \right\rbrace \right]
\nonumber
\\
&+\left(\frac{n-4}{n-3}\right)\kappa _{n}^{4}\left[\Pi _{\alpha \beta}e^{\alpha}_{A}e^{\beta}_{B}
+q_{AB}\left\lbrace \Pi _{\alpha \beta}s^{\alpha}s^{\beta}-\frac{1}{n-2}\Pi \right\rbrace \right]
\nonumber
\\
&-\left(\frac{n-4}{n-3}\right)\left[E_{\alpha \beta}e^{\alpha}_{A}e^{\beta}_{B}
+q_{AB}\left\lbrace E_{\alpha \beta}s^{\alpha}s^{\beta}-\frac{E}{n-2} \right\rbrace \right]
-\mathcal{E}_{AB}
\nonumber
\\
&+\kappa _{n-1}^{4}\Upsilon _{AB}
\nonumber
\\
&-\kappa _{n-1}^{4}\Sigma ^{2}\frac{(n-4)}{8(n-3)}q_{AB}
+\kappa _{n-1}^{4}\Sigma \frac{(n-4)}{4(n-3)}\nu_{AB}
\end{align}
where, we have introduced, the following quantities,
\begin{align}
\Pi _{\alpha \beta}&=\left[-\frac{1}{4}t_{\alpha \mu}t^{\mu}_{\beta}
+\frac{1}{8}h_{\alpha \beta}t_{\mu \nu}t^{\mu \nu}
+\frac{1}{4(n-2)}t t_{\alpha \beta}
-\frac{1}{8(n-2)}h_{\alpha \beta}t^{2}\right]
\\
\Upsilon _{AB} &=\left[-\frac{1}{4}\nu _{AC}\nu ^{C}_{B}+\frac{1}{8}q_{AB}\nu _{CD}\nu ^{CD}
+\frac{1}{4(n-3)}\nu \nu _{AB}-\frac{1}{8(n-3)}q_{AB}\nu ^{2} \right]
\end{align}
The effective equation can be written in conventional form after some algebraic manipulations such that:
\begin{align}
^{n-1}G_{AB}=\kappa _{n-2}^{2}T_{AB}
\end{align}
where energy momentum tensor $T_{AB}$ has the expression:
\begin{align}
T_{AB}&=-\Lambda _{n-2}q_{AB}+\nu _{AB}+\frac{\kappa _{n-1}^{4}}{\kappa _{n-2}^{2}}\Upsilon _{AB}
-\frac{1}{\kappa _{n-2}^{2}}\mathcal{E}_{AB}
\nonumber
\\
&+\left(\frac{n-4}{n-3}\right)\frac{\kappa _{n-1}^{2}}{\kappa _{n-2}^{2}}\left[t_{\alpha \beta}e^{\alpha}_{A}e^{\beta}_{B}
+q_{AB}\left\lbrace t_{\alpha \beta}s^{\alpha}s^{\beta}-\frac{1}{n-2}t \right\rbrace \right]
\nonumber
\\
&+\left(\frac{n-4}{n-3}\right)\frac{\kappa _{n}^{4}}{\kappa _{n-2}^{2}}\left[\Pi _{\alpha \beta}e^{\alpha}_{A}e^{\beta}_{B}
+q_{AB}\left\lbrace \Pi _{\alpha \beta}s^{\alpha}s^{\beta}-\frac{1}{n-2}\Pi \right\rbrace \right]
\nonumber
\\
&-\left(\frac{n-4}{n-3}\right)\frac{1}{\kappa _{n-2}^{2}}\left[E_{\alpha \beta}e^{\alpha}_{A}e^{\beta}_{B}
+q_{AB}\left\lbrace E_{\alpha \beta}s^{\alpha}s^{\beta}-\frac{E}{n-2} \right\rbrace \right]
\end{align}
Here, we have introduced the following two objects for notational simplicity,
\begin{align}
\Lambda _{n-2}&=\left(\frac{n-4}{n-2}\right)\frac{\kappa _{n-1}^{2}}{\kappa _{n-2}^{2}}\left[\Lambda _{n-1}
+\frac{(n-2)}{8(n-3)}\kappa _{n-1}^{2}\Sigma ^{2} \right]
\\
\kappa _{n-2}^{2}=8\pi G_{n-2}&=\kappa _{n-1}^{4}\Sigma \frac{(n-4)}{4(n-3)}
\end{align}
where $\Lambda _{n-2}$ and $\kappa _{n-2}$ can be interpreted as the $(n-2)$-dimensional cosmological constant and $(n-2)$ dimensional gravitational constant respectively. However for the case, in which $t_{\alpha \beta}=0$, i.e., the bulk energy momentum tensor vanishes, the above equation gets simplified to,
\begin{align}
^{(n-2)}G_{AB}=\kappa _{n-2}^{2}T_{AB}
\end{align}
where $T_{AB}$ has the simplified form:
\begin{align}
T_{AB}&=-\Lambda _{n-2}q_{AB}+\nu _{AB}+\frac{\kappa _{n-1}^{4}}{\kappa _{n-2}^{2}}\Upsilon _{AB}
-\frac{1}{\kappa _{n-2}^{2}}\mathcal{E}_{AB}
\nonumber
\\
&-\left(\frac{n-4}{n-3}\right)\frac{1}{\kappa _{n-2}^{2}}\left[E_{\alpha \beta}e^{\alpha}_{A}e^{\beta}_{B}
+q_{AB}\left\lbrace E_{\alpha \beta}s^{\alpha}s^{\beta}-\frac{E}{n-2} \right\rbrace \right]
\end{align}
Thus the structure of effective field equation in $(n-2)$-dimensional brane for vanishing bulk energy momentum tensor is mostly identical to the $(n-1)$-dimensional counterpart except for the Weyl curvature parts. The structure of the effective equation should be quiet evident at this stage, which we are going to exploit later to obtain effective field equation in $(n-m)$-dimensional brane.
\subsection{Generalization to (n-m)-dimensional Brane}\label{Sec03:n-m}

Having discussed the effective gravitational field equation in $(n-2)$-dimension starting from $n$-dimension through a two step decomposition, it is now time to consider the generalization of the effective field equation for gravity to $(n-m)$-dimension. This can be obtained following the analogy of previous results. The final expression for effective field equation in $(n-m)$ dimension is:
\begin{align}\label{GCNew01}
^{(n-m)}G_{AB}=\kappa _{n-m}^{2}T_{AB}^{(n-m)}
\end{align}
where the energy momentum tensor has the following expression:
\begin{align}\label{GCNew02}
T_{AB}^{(n-m)}&=-\Lambda _{n-m}q_{AB}^{(n-m)}+\tau _{AB}^{(n-m)}+\frac{\kappa _{n-m+1}^{4}}{\kappa _{n-m}^{2}}\Upsilon _{AB}^{(n-m)}-\frac{1}{\kappa _{n-m}^{2}}E_{AB}^{(n-m)}
\nonumber
\\
&+\sum _{i=2}^{m}\left(\frac{n-i-2}{n-i-1}\right)\Big[\frac{\kappa _{n-i+1}^{2}}{\kappa _{n-i}^{2}} \left(t_{\alpha \beta}^{(n-i+1)}e^{\alpha}_{A}e^{\beta}_{B}
+q_{AB}\left\lbrace t_{\alpha \beta}^{(n-i+1)}s^{\alpha}s^{\beta}-\frac{1}{n-2}t^{(n-i+1)} \right\rbrace \right)
\nonumber
\\
&+\frac{\kappa _{n-i+2}^{4}}{\kappa _{n-i}^{2}}\left(\Pi _{\alpha \beta}^{(n-i+1)}e^{\alpha}_{A}e^{\beta}_{B}
+q_{AB}\left\lbrace \Pi _{\alpha \beta}^{(n-i+1)}s^{\alpha}s^{\beta}-\frac{1}{n-2}\Pi ^{(n-i+1)} \right\rbrace \right)
\nonumber
\\
&-\frac{1}{\kappa _{n-i}^{2}} \left(E_{\alpha \beta}^{(n-i+1)}e^{\alpha}_{A}e^{\beta}_{B}
+q_{AB}\left\lbrace E_{\alpha \beta}^{(n-i+1)}s^{\alpha}s^{\beta}-\frac{E^{(n-i+1)}}{n-2} \right\rbrace \right)\Big]
\end{align}
with the following identifications:
\begin{align}
\Lambda _{n-m}&=\left(\frac{n-m-2}{n-m}\right)\frac{\kappa _{n-m+1}^{2}}{\kappa _{n-m}^{2}}\left(\Lambda _{n-m+1}+\frac{(n-m)}{8(n-m-1)}\kappa _{n-m+1}\Sigma ^{2}\right)
\\
\kappa _{n-m}^{2}&=8\pi G_{n-m}=\frac{(n-m-2)}{4(n-m-1)}\Sigma ^{(n-m)}\kappa _{n-m+1}^{4}
\\
\Pi _{\alpha \beta}&=\left[-\frac{1}{4}t_{\alpha \mu}t^{\mu}_{\beta}
+\frac{1}{8}h_{\alpha \beta}t_{\mu \nu}t^{\mu \nu}
+\frac{1}{4(n-m)}t t_{\alpha \beta}
-\frac{1}{8(n-m)}h_{\alpha \beta}t^{2}\right]
\\
\Upsilon _{AB} &=\left[-\frac{1}{4}\tau _{AC}\tau ^{C}_{B}+\frac{1}{8}q_{AB}\tau _{CD}\tau ^{CD}
+\frac{1}{4(n-m-1)}\tau \tau _{AB}-\frac{1}{8(n-m-1)}q_{AB}\tau ^{2} \right]
\end{align}
However it may be observed that in the situation where there is only matter on the $(n-m)$-dimensional brane but not on the other higher dimensional branes, all the additional terms vanishes except the Weyl tensor part. If the bulk spacetime is assumed to be Anti de-Sitter (AdS) then the Weyl tensor can taken to be zero, in which case the effective field equation takes a simplified form:
\begin{align}
^{(n-m)}G_{AB}=\kappa _{n-m}^{2}\left(-\Lambda _{n-m}q_{AB}^{(n-m)}+\tau _{AB}^{(n-m)}+\frac{\kappa _{n-m+1}^{4}}{\kappa _{n-m}^{2}}\Upsilon _{AB}^{(n-m)}\right)
\end{align}
Thus in this simple situation the effective field equation has contribution from $(n-m)$ dimensional cosmological constant, energy momentum tensor on the brane and higher order correction terms originating from $\Upsilon _{AB}$. Having derived the effective field equation in Einstein gravity let us now consider the identical situation in $f(\mathcal{R})$ gravity as well.
\section{Applications in General Relativity}\label{AppGR}

Having derived the effective field equation on an arbitrary lower dimensional hypersurface starting from a n-dimensional bulk, in this section we will consider two applications of our result. One of these applications would be for static, spherically symmetric configuration and the other in cosmology. As an illustration we will take the bulk to be six dimensional while the brane is obtained by two step decomposition and is four dimensional. 
\subsection{Static Spherically Symmetric Vacuum Brane}

We start by considering the case for static, spherically symmetric spacetime with no matter source present. Thus brane energy momentum tensor vanishes however the electric part of the Weyl tensor survives. Hence the effective gravitational field equation on the four dimensional brane starting from the six dimensional bulk can be obtained from \eqs{GCNew01} and (\ref{GCNew02}) as,
\begin{align}
^{(4)}G_{AB}=-\mathcal{E}_{AB}-\frac{2}{3}\left[E_{\alpha \beta}e^{\alpha}_{A}e^{\beta}_{B}+q_{AB}\left(E_{\alpha \beta}s^{\alpha}s^{\beta}-\frac{1}{4}E\right)\right]
\end{align}
In arriving at the above equation the cosmological constant has been neglected, since we are interested in a local spacetime region. The electric part of the Weyl tensor in a static, spherically symmetric spacetime can be written using two unknown functions depending on the radial coordinate as,
\begin{align}
\mathcal{E}_{AB}=-\frac{\kappa _{5}^{4}}{\kappa _{4}^{4}}\left[U(r)\left(u_{A}u_{B}+\frac{1}{3}\bar{q}_{AB}\right)+P(r)\left(r_{A}r_{B}-\frac{1}{3}\bar{q}_{AB}\right)\right]
\end{align}
where $\bar{q}_{AB}$ is the induced metric on the $t=\textrm{constant}$ surface, such that $u^{A}\bar{q}_{AB}=0$. The five dimensional electric part of Weyl tensor would have similar decomposition depending on two separate radial functions $\bar{U}(r)$ and $\bar{P}(r)$. On projecting this five dimensional tensor $E_{\alpha \beta}$ on the four dimensional brane, the effective field equations reduce to
\begin{align}
G_{AB}&=\frac{\kappa _{5}^{4}}{\kappa _{4}^{4}}\left[U(r)\left(u_{A}u_{B}+\frac{1}{3}\bar{q}_{AB}\right)+P(r)\left(r_{A}r_{B}-\frac{1}{3}\bar{q}_{AB}\right)\right]
\nonumber
\\
&+\frac{2}{3}\frac{\kappa _{6}^{4}}{\kappa _{5}^{4}}\left[\bar{U}(r)\left(u_{A}u_{B}+\frac{1}{2}\bar{q}_{AB}\right)+\bar{P}(r)\left(r_{A}r_{B}-\frac{1}{2}\bar{q}_{AB}\right)\right]
\end{align}
Now $\kappa _{n}$ and $\kappa _{n-1}$ are related through \eq{GC14}, such that, $\kappa _{6}^{4}=(16/3)\kappa _{5}^{2}(1/\sigma)$, $\kappa _{5}^{4}=(6/\Sigma)\kappa _{4}^{2}$ and $\kappa _{4}^{2}=8\pi G$. Using these relations the above equation can be written in terms of the brane tensions, $\sigma$ for the five dimensional brane, $\Sigma$ for the four dimensional brane and the Gravitational constant $G$ as,
\begin{align}
G_{AB}&=\frac{3}{4\pi G \Sigma}\left[U(r)\left(u_{A}u_{B}+\frac{1}{3}\bar{q}_{AB}\right)+P(r)\left(r_{A}r_{B}-\frac{1}{3}\bar{q}_{AB}\right)\right]
\nonumber
\\
&+\frac{32}{9\sqrt{6}}\frac{\sqrt{\Sigma}}{\sigma \sqrt{8\pi G}}\left[\bar{U}(r)\left(u_{A}u_{B}+\frac{1}{2}\bar{q}_{AB}\right)+\bar{P}(r)\left(r_{A}r_{B}-\frac{1}{2}\bar{q}_{AB}\right)\right]
\end{align}
Now assuming a static, spherically symmetric metric ansatz of the form,
\begin{align}
ds^{2}=-e^{\nu (r)}dt^{2}+e^{\lambda (r)}dr^{2}+r^{2}d\Omega ^{2}
\end{align}
we obtain various components of the gravitational field equations to be
\begin{align}
-e^{-\lambda}\left(-\frac{\lambda '}{r}+\frac{1}{r^{2}}\right)+\frac{1}{r^{2}}&=\frac{3}{4\pi G \Sigma}U(r)+\frac{32}{9\sqrt{6}}\frac{\sqrt{\Sigma}}{\sigma \sqrt{8\pi G}}\bar{U}(r)
\label{GCEqNew01}
\\
e^{-\lambda}\left(\frac{\nu '}{r}+\frac{1}{r^{2}}\right)-\frac{1}{r^{2}}&=\frac{1}{4\pi G \Sigma}\left[U(r)+2P(r)\right]+\frac{16}{9\sqrt{6}}\frac{\sqrt{\Sigma}}{\sigma \sqrt{8\pi G}}\left[\bar{U}(r)+\bar{P}(r)\right]
\label{GCEqNew02}
\\
\frac{1}{2}e^{-\lambda}\left(\nu ''+\frac{\nu '^{2}}{2}+\frac{\nu '-\lambda '}{r}-\frac{\nu '\lambda '}{2}\right)&=\frac{1}{4\pi G \Sigma}\left[U(r)-P(r)\right]+\frac{16}{9\sqrt{6}}\frac{\sqrt{\Sigma}}{\sigma \sqrt{8\pi G}}\left[\bar{U}(r)-\bar{P}(r)\right]
\end{align}
Hence the effect of multiple extra dimensions is manifested in the lower dimensional effective gravitational field equations through the appearance of extra terms $\bar{U}(r)$ and $\bar{P}(r)$. Field equations have to be supplemented by the conservation equations, which will provide a differential relation connecting $U(r)$, $P(r)$, $\bar{U}(r)$ and $\bar{P}(r)$. The conservation equations turn out to be,
\begin{align}
\frac{1}{3}\left(\frac{dU}{dr}+2\frac{dP}{dr}\right)+\frac{\nu '}{3}\left(2U+P\right)+\frac{2P}{r}&=0
\label{GC_Conserv01}
\\
\frac{1}{2}\left(\frac{d\bar{U}}{dr}+\frac{d\bar{P}}{dr}\right)+\frac{\nu '}{4}\left(3\bar{U}+\bar{P}\right)+\frac{2\bar{P}}{r}&=0
\label{GC_Conserv02}
\end{align}
Starting from \eq{GCEqNew01} it is straightforward to obtain a solution for $e^{-\lambda}$. Since all the objects on the right hand side of \eq{GCEqNew01} depends only on the radial coordinate we can integrate them to obtain the following solution for $e^{-\lambda}$ as,
\begin{align}\label{New01}
e^{-\lambda}=1-\frac{2GM}{r}-\frac{Q(r)}{r};\qquad Q(r)=\frac{3}{4\pi G \Sigma}\int dr ~r^{2}U(r)+\frac{32}{9\sqrt{6}}\frac{\sqrt{\Sigma}}{\sigma \sqrt{8\pi G}}\int dr ~r^{2}\bar{U}(r)
\end{align}
Having obtained $e^{-\lambda}$ it is now pertinent to consider the other metric component $e^{\nu}$ as well. This involves all the extra contributions from dark radiation and dark pressure terms. In order to obtain solutions for these higher dimensional contributions, we need to impose certain relations among these higher dimensional objects. The most useful among them being the relations $2U+P=0$ and $3\bar{U}+\bar{P}=0$. Then from \eq{GC_Conserv01} and \eq{GC_Conserv02} we readily obtain,
\begin{align}
P(r)&=-2U(r)=\frac{P_{0}}{r^{4}};\qquad \bar{P}(r)=-3\bar{U}(r)=\frac{\bar{P}_{0}}{r^{6}}
\\
Q(r)&=Q_{0}+\frac{3\alpha P_{0}}{2r}+\frac{\bar{P}_{0}\beta}{9r^{3}};\qquad \alpha =\frac{1}{4\pi G\Sigma};\qquad \beta =\frac{32}{9\sqrt{6}}\frac{\sqrt{\Sigma}}{\sigma \sqrt{8\pi G}}
\end{align}
For these relations between dark radiation and dark pressure the static spherically symmetric solution corresponds to, 
\begin{align}\label{GC_Sol01}
e^{-\lambda}=e^{\nu}=1-\frac{2GM+Q_{0}}{r}-\frac{3\alpha P_{0}}{2r^{2}}-\frac{\bar{P}_{0}\beta}{9r^{4}}
\end{align}
Note that this is an exact solution to the effective field equations, no assumptions or approximations have been made. Hence the effect of extra spatial dimension appears through the charge term, i.e., $r^{-2}$ term and the $r^{-4}$ term. Thus near $r=0$, $1/r^{4}$ term would dominate, which originates from the existence of more than one extra spatial dimensions.

Having derived the static spherically symmetric solution there are two important aspects one can study --- (a) stability of the solutions and (b) thermodynamics. We will briefly comment on these features. The stability of solution can be obtained by considering perturbations, which again are of three types, scalar, vector and tensor modes. These modes satisfy a set of decoupled wave equations with the structure,
\begin{align}
\left(\square -\frac{1}{f(r)}V\right)\Phi =0
\end{align}
Here $\square$ represents d'Alembertian operator, $f(r)=e^{\nu}=e^{-\lambda}$, $\Phi$ stands for the perturbation modes and $V$ is an appropriate potential associated with perturbation $\Phi$. For stability, the potentials should be positive. If we assume that the corrections to the Schwarzschild solution is small, i.e., $\alpha$ and $\beta$ are small compared to $M$, then the above solution is stable for all the modes.

Let us now discuss the case of black hole thermodynamics. For the horizon location we have the following equation,
\begin{align}
1-\frac{2GM+Q_{0}}{r_{h}}-\frac{3\alpha P_{0}}{2r_{h}^{2}}-\frac{\bar{P}_{0}\beta}{9r_{h}^{4}}=0
\end{align}
The mass of the black hole (or, equivalently the internal energy) can be obtained readily from the above equation as,
\begin{align}\label{Eq01}
M(r_{h})=\frac{r_{h}}{2}-\frac{Q_{0}}{2}-\frac{3\alpha P_{0}}{4r_{h}}-\frac{\bar{P}_{0}\beta}{6r_{h}^{3}}
\end{align}
Now entropy can be obtained from holography, i.e., $S=A/4$, such that, $S=r_{h}^{2}$. Thus internal energy can be obtained in terms of entropy density from \eq{Eq01} such that,
\begin{align}
M(S)=\frac{\sqrt{S}}{2}-\frac{Q_{0}}{2}-\frac{3\alpha P_{0}}{4\sqrt{S}}-\frac{\bar{P}_{0}\beta}{6S^{3/2}}
\end{align}
Then the black hole temperature can be obtained in terms of entropy as,
\begin{align}
T(S)=\left(\frac{\partial M}{\partial S}\right)=\frac{1}{4\sqrt{S}}+\frac{3\alpha P_{0}}{8S^{3/2}}+\frac{\bar{P}_{0}\beta}{4S^{5/2}}
\end{align}
which immediately leads to the following expression for specific heat as,
\begin{align}\label{cvexp}
C_{V}=T\left(\frac{\partial S}{\partial T}\right)=-\frac{\frac{1}{4\sqrt{S}}+\frac{3\alpha P_{0}}{8S^{3/2}}+\frac{\bar{P}_{0}\beta}{4S^{5/2}}}{\frac{1}{8S^{3/2}}+\frac{9\alpha P_{0}}{16S^{5/2}}+\frac{5\bar{P}_{0}\beta}{8S^{7/2}}}
\end{align}
Thus the specific heat of the black hole is negative, as it should and there is no phase transition possible, since $\alpha$ and $\beta$ are both positive quantities. Later we will show that inclusion of $f(\mathcal{R})$ gravity can alter the situation drastically, which we will discuss in the next section.

Having derived the static spherically symmetric solution and its various properties, let us now discuss some novel features of this solution
\begin{itemize}

\item Imposing the following restrictions $2U(r)+P(r)=0$ and $\bar{P}(r)+3\bar{U}(r)=0$, we have shown that the metric elements satisfy $e^{-\lambda}=e^{\nu}$. Thus the line element derived in \eq{GC_Sol01} depends explicitly on the form of $U(r)$ and $\bar{U}(r)$. Thus the geometry of the spacetime is uniquely determined only when knowledge about multiple dimensions are obtained through $\bar{U}(r)$. Also it turns out that at small $r$ the effect from $\bar{U}(r)$ dominates.

\item The solution is also stable under perturbations. The potentials associated with respective wave equations are positive for scalar, vector and tensor modes. Thus the black hole solution, will not change its character under perturbations.

\item Finally, the specific heat associated with the black hole solution is negative. This shows that the black hole is thermodynamically unstable. Since all the additional parameters are positive there is no phase transition in this black hole system.

\end{itemize}
Hence the fact that spacetime has more than one extra dimensions is being reflected on the right hand side of the effective gravitational field equations. These extra terms contribute to the energy momentum tensor and modifies the metric components from just being $1-(2GM/r)$. 
\subsection{Cosmological Spacetime}

After discussing the static spherically symmetric configuration it is natural to turn around and consider gravity theory on a lower dimensional hypersurface in a cosmological context. As already mentioned we will start with $6$-dimensional bulk spacetime and then obtain cosmological solutions from the effective field equation on a 4-dimensional brane. In order to obtain analytical solution, we assume $E_{ab}=0$ i.e. contribution from electric part of Weyl tensor vanishes. Then effective field equation in 4-dimension takes the following form:
\begin{align}
G_{AB}&=-\kappa _{4}^{2}\Lambda _{4}q_{AB}+\kappa _{4}^{2}\tau _{AB}+\kappa _{5}^{4}\Upsilon _{AB}
\nonumber
\\
&+\frac{2}{3}\kappa _{5}^{2}\left(t_{\alpha \beta}e^{\alpha}_{A}e^{\beta}_{B}
+q_{AB}\left\lbrace t_{\alpha \beta}s^{\alpha}s^{\beta}-\frac{1}{4}t\right\rbrace \right)
\nonumber
\\
&+\frac{2}{3}\kappa _{6}^{4}\left(\Pi _{\alpha \beta}e^{\alpha}_{A}e^{\beta}_{B}
+q_{AB}\left\lbrace \Pi _{\alpha \beta}s^{\alpha}s^{\beta}-\frac{1}{4}\Pi \right\rbrace \right)
\end{align} 
In this expression along with brane energy momentum tensor $\tau _{AB}$ and its higher order term $\Upsilon _{AB}$, two additional contribution from 5-dimensional matter energy-momentum tensor is present. We assume that the brane is filled with perfect fluid with the following form for the energy-momentum tensor on the 4-dimension as:
\begin{align}
\tau ^{A}_{B}=\textrm{diag}\left(-\rho ,p,p,p\right)
\end{align}
where $\rho$ represents energy density of the perfect fluid and $p$ yields its pressure. The energy-momentum tensor in 5-dimension is taken to be pressure free dust, such that its energy momentum tensor has the simple form:
\begin{align}
t^{\alpha}_{\beta}=\textrm{diag}\left(-\rho _{5},0,0,0,0\right)
\end{align}
where the subscript `5' indicates that it is from 5-dimensional brane. Using which we arrive at the following expression for the higher order components
\begin{align}
\Pi _{tt}=\frac{3}{32}\rho _{5}^{2};\qquad \Pi =-\frac{3}{32}\rho _{5}^{2}
\end{align}
Having obtained all the components of the matter energy momentum tensor, the time-time and spatial components of effective equations lead to (we will assume that the spacetime is spatially flat, i.e., $k=0$):
\begin{align}
3H^{2}&=\kappa _{4}^{2}\Lambda _{4}+\kappa _{4}^{2}\rho +\frac{1}{12}\kappa _{5}^{4}\rho ^{2}+\frac{1}{2}\kappa _{5}^{2}\rho _{5}+\frac{3}{64}\kappa _{6}^{4}\rho _{5}^{2}
\label{FRW01a}
\\
H^{2}+2\frac{\ddot{a}}{a}&=\kappa _{4}^{2}\Lambda _{4}-\kappa _{4}^{2}p -\frac{1}{12}\kappa _{5}^{4}\left(\rho ^{2}+2p\rho \right)-\frac{1}{6}\kappa _{5}^{2}\rho _{5}-\frac{1}{64}\kappa _{6}^{4}\rho _{5}^{2}
\label{FRW01b}
\end{align}
Having obtained the Friedmann equations, let us explore the various phases of the universe. To get a handle, let us write the above expression in terms of relative abundances $\Omega _{i}=\rho _{i,0}/\rho _{c}$. Here $\rho _{i,0}$ stands for the present energy density of the ith component and $\rho _{c}$ is the current total energy density of the universe. The energy density can be expanded in terms of baryonic and relativistic matter fields. We will assume that matter in higher dimension to be non-relativistic. Thus the Hubble parameter turns out to be,
\begin{align}
H^{2}=H_{0}^{2}\left[\Omega _{\Lambda}+\left(\Omega _{B}+\Omega _{5}\right)\left(\frac{a_{0}}{a}\right)^{3}+\Omega _{R}\left(\frac{a_{0}}{a}\right)^{4}\right]
\end{align}
where, $\Omega _{5}=(\kappa _{5}^{2}/16\pi G \ell)(\rho _{5,0}/\rho _{c})$, with $\ell$ being a characteristic length associated with the extra dimension. Note that the above Hubble parameter can reproduce the standard $\Lambda CDM$ model. For at late times $\Omega _{\Lambda}$ would dominate leading to accelerated expansion of the universe. At early times, the $1/a^{4}$ term would dominate, leading to radiation dominated universe. The intermediate region is dominated by non-relativistic matter such that $H^{2}\sim a^{-3}$. However the surprising fact about this model is that in the matter sector along with the baryonic matter, due to extra dimension we have additional contribution, which can act as dark matter. In standard $\Lambda CDM$ cosmology, the dark matter has to be included in the model by hand. However in this case it appears in the Friedmann universe \emph{naturally}, we need not have to invoke it by hand. Using the epoch of matter, radiation equality it is possible to get a numerical estimate 
about five-dimensional gravitational constant as,
\begin{align}
1+z_{eq}=3.9\times 10^{4}\left(0.022+\frac{\kappa _{5}^{2}h^{2}}{16\pi G \ell}\frac{\rho _{5,0}}{\rho _{c}}\right)
\end{align}
where $h=0.72$, $\ell$ stands for the length of the extra dimension. Thus knowing the redshift $z_{eq}$, we can determine the value of the higher dimensional gravitational constant $\kappa _{5}^{2}$. Thus spacetime having more than one extra spatial dimensions can naturally provide additional matter components in the right hand side of the Friedmann equations. To have inflation we need to include an additional scalar field on top of this alike the standard $\Lambda CDM$ model. Thus the effective field equations are capable to explain the phases of the universe.

Writing $\kappa _{4}^{2}\Lambda _{4}=\Lambda _{\textrm{eff}}$ and for the bulk matter we have $\rho _{5}=(\rho _{5,0}/\ell)(1/a^{3})$, where $\ell$ represents the finite length of the 5-dimensional brane coordinate. Assuming that the matter in the 4-dimensional brane is also pressure-less, we have the following expression for the Hubble parameter: 
\begin{align}
H^{2}=\frac{\Lambda _{\textrm{eff}}}{3}+\frac{8\pi G}{3}\frac{\rho _{0}}{a^{3}} +\frac{1}{36}\kappa _{5}^{4}\frac{\rho _{0}^{2}}{a^{6}}+\frac{1}{6}\kappa _{5}^{2}\frac{\rho _{5,0}}{\ell}\frac{1}{a^{3}}+\frac{1}{64}\kappa _{6}^{4}\frac{\rho _{5,0}^{2}}{\ell ^{2}}\frac{1}{a^{6}}
\end{align}
where $\rho _{0}$ represents the value of energy density in the present epoch and $\rho _{5,0}$ represents the same for 5-dimensional matter. The above expression for Hubble parameter can be simplified and arranged properly leading to:
\begin{align}\label{HubbleMultipleNew}
H^{2}=\frac{\Lambda _{\textrm{eff}}}{3}+\left\lbrace \frac{8\pi G}{3}\rho _{0}+\frac{1}{6}\kappa _{5}^{2}\frac{\rho _{5,0}}{\ell} \right\rbrace \frac{1}{a^{3}}+\left\lbrace \frac{1}{36}\kappa _{5}^{4}\rho _{0}^{2}+\frac{1}{64}\kappa _{6}^{4}\frac{\rho _{5,0}^{2}}{\ell ^{2}}\right\rbrace \frac{1}{a^{6}}
\end{align}
Then following the techniques developed in \cite{Chakraborty2014d} we can introduce three constants, namely:
\begin{align}
C_{\Lambda}=\frac{\Lambda _{\textrm{eff}}}{3};\qquad C_{\rho}=\frac{8\pi G}{3}\rho _{0}+\frac{1}{6}\kappa _{5}^{2}\frac{\rho _{5,0}}{\ell};\qquad C_{\textrm{Sq}}=\frac{1}{36}\kappa _{5}^{4}\rho _{0}^{2}+\frac{1}{64}\kappa _{6}^{4}\frac{\rho _{5,0}^{2}}{\ell ^{2}}
\end{align}
such that we obtain the following solution to the scale factor as:
\begin{align}\label{GC72New}
\exp \left[3\sqrt{C_{\Lambda}}\left(t-t_{0}\right)\right]&=
\frac{2\sqrt{C_{\Lambda}}\sqrt{C_{\Lambda}a^{6}+C_{\rho}a^{3}+C_{\textrm{Sq}}}+2C_{\Lambda}a^{3}+C_{\rho}}
{2\sqrt{C_{\Lambda}}\sqrt{C_{\Lambda}+C_{\rho}+C_{\textrm{Sq}}}+2C_{\Lambda}+C_{\rho}};\qquad C_{\Lambda}>0
\\
3\sqrt{-C_{\Lambda}}\left(t_{0}-t\right)&=
\sin ^{-1} \left(\frac{2C_{\Lambda}a^{3}+C_{\rho}}{\sqrt{C_{\rho}^{2}-4C_{\Lambda}C_{\textrm{Sq}}}}\right)
-\sin ^{-1} \left(\frac{2C_{\Lambda}+C_{\rho}}{\sqrt{C_{\rho}^{2}-4C_{\Lambda}C_{\textrm{Sq}}}} \right);
\nonumber
\\
\qquad C_{\Lambda}&<0,\qquad C_{\rho}^{2}>4C_{\Lambda}C_{\textrm{Sq}}
\end{align}
Though the solutions have a complicated appearance, we can impose certain conditions under which the solutions simplify substantially, yielding clearer physical insight. For that purpose we consider the situation $\Lambda _{\textrm{eff}}=0$. This would be true if we assume the 5-dimensional cosmological constant to be negative, such that: $\Lambda _{5}=-(\kappa _{5}^{2}\Sigma ^{2})/6$. Under this condition the Hubble parameter from \eq{HubbleMultipleNew} turns out to be,
\begin{equation}
H^{2}=\frac{8\pi G}{3}\rho \left[1+\frac{\kappa _{5}^{2}}{16\pi G}\frac{\rho _{5}}{\rho}+\frac{\rho}{2\Sigma}
\left(1+\frac{9}{16}\frac{\kappa _{6}^{4}}{\kappa _{5}^{4}}\frac{\rho _{5}^{2}}{\rho ^{2}}\right)\right]
\end{equation}  
If we assume that $\rho$ represents energy density of non-relativistic matter we arrive at the following differential equation
\begin{equation}
H^{2}=\left\lbrace \frac{8\pi G \rho _{0}}{3}+\frac{\kappa _{5}^{2}}{6}\frac{\rho _{5,0}}{\ell}\right\rbrace \frac{1}{a^{3}}+\frac{4\pi G \rho _{0}^{2}}{3\Sigma}
\left(1+\frac{9}{16}\frac{\kappa _{6}^{4}}{\kappa _{5}^{4}}\frac{\rho _{5,0}^{2}}{\ell ^{2}\rho _{0} ^{2}}\right)\frac{1}{a^{6}}
\end{equation}
leading to the following solution for the scale factor:
\begin{equation}\label{New02}
a^{3}=\left\lbrace 6\pi G \rho _{0}+3\frac{\kappa _{5}^{2}}{8}\frac{\rho _{5,0}}{\ell}\right\rbrace t^{2}+\left\lbrace \sqrt{\frac{12\pi G \rho _{0}^{2}}{\Sigma}
\left(1+\frac{9}{16}\frac{\kappa _{6}^{4}}{\kappa _{5}^{4}}\frac{\rho _{5,0}^{2}}{\ell ^{2}\rho _{0} ^{2}}\right)}\right\rbrace t
\end{equation}
It is clear from the above expression that the universe undergoes a transition in the expansion rate. In the standard scenario there are three phases, the radiation dominated, the matter dominated and the accelerated expansion. If we treat the additional correction terms due to existence of extra spatial dimensions with importance then even within matter dominated epoch we will have transition. The time scale when this happened depends on whether the brane matter energy dominates over the bulk matter or not. Below we present the result for timescale of transition for both the situations:  
\begin{align}
t\sim \sqrt{\frac{1}{3\pi G \Sigma}}&=\frac{4}{\Sigma \kappa _{5}^{2}}
=-\frac{2}{3}\Lambda _{5}^{-1};
\qquad (\rho _{5,0}/\ell \ll \rho _{0})
\\
t \sim & \frac{\kappa _{6}^{2}}{\kappa _{5}^{2}}\left(1-\frac{16\pi G\rho _{0}\ell}{\kappa _{5}^{2}\rho _{5,0}}\right)
\qquad (\rho _{5,0}/\ell \gg \rho _{0})
\end{align}
Thus at early universe (within the matter dominated epoch) we have a high energy regime, where $a\sim t^{1/3}$, while at late time low energy regime the scale factor variation with time modifies to $a\sim t^{2/3}$, which is the standard evolution of the matter field. Note that the same should happen in radiation dominated epoch as well. Hence the scale factor at later times will behave as it should in a standard $\Lambda CDM$ model. In a nutshell, the effect of extra dimensions on cosmological scenario can be seen from the following results:
\begin{itemize}

\item It is clear from the expression for Hubble parameter that the 5-dimensional matter adds to the 4-dimensional one and \emph{enhances} the total non-relativistic matter content of the universe (4-dimensional brane we live in). Hence matter fields present in the extra dimension can act as a \emph{viable} source of dark matter. The most important part of the above analysis corresponds to the fact that the dark matter contribution appears \emph{naturally}, we need not have to add them by hand (a related issue has also been addressed in \cite{Chakraborty2007}).

\item The existence of more than one extra spatial dimension introduces additional correction terms in the Friedmann equations. This results in transition within both matter dominated and radiation dominated epoch. This in principle can affect the structure formation in the universe and hence can be used to constrain various higher dimensional parameters.

\end{itemize}
Thus with matter present in the 5-dimensional brane we can have standard cosmology with proper scaling of the scale factor with time. However the matter content and transition time scale will depend on the content in the extra dimensions. 

The cosmological solutions in brane models have interesting holographic interpretations. It was shown in \cite{Verlinde2000} that the usual four dimensional cosmological equations can be re-written in an equivalent form of the entropy of two dimensional conformal field theory (CFT). In our case the Friedman equations are given by \eqs{FRW01a} and (\ref{FRW01b}) respectively. Now defining the Hawking, Bekenstein and Bekenstein-Hawking entropies as,
\begin{align}
S_{H}=\frac{(n-1)HV}{4G},\qquad S_{BH}=\frac{(n-1)V}{4Ga},\qquad S_{B}=\frac{2\pi a}{n}E
\end{align}
where $E=\rho V$ is the total energy within a volume V and $\rho _{m}\propto a^{-n}$. Also the total entropy should be bounded by the Hubble entropy $S_{H}$, which is the entropy of the black hole with the radius of the Hubble size. Using these definitions we can rewrite the Friedman equations in the form,
\begin{align}
S_{H}=\frac{2\pi a}{n}\sqrt{E_{BH}\left(2E-KE_{BH}\right)};\qquad KE_{BH}=n\left(E+pV-T_{H}S_{H}\right)
\end{align}
with energy of the black hole being $E_{BH}=n(n-1)V/8\pi Ga^{2}$ and Hawking temperature being $T_{H}=-\dot{H}/2\pi H$. This is exactly the form one would have from the CV entropy relation. It explicitly shows the connection between four dimensional classical gravitational physics and 2-dimensional quantum 2-dimensional CFT \cite{Cardy1986}. This two dimensional CFT entropy representation of cosmological equations is called generalized Cardy or Cardy-Verlinde (CV) formula. One can also consider dS or AdS black holes in the bulk, such that the brane equations of motion can be casted in the form of Friedman equations. By introducing brane matter the cosmological equations can be written in the CV form which is related to AdS black hole entropy (these and various other aspects have been discussed in \cite{Nojiri2002} quiet extensively). Now we will generalize these results to $f(\mathcal{R})$ gravity model.
\section{Generalization to bulk f(R) gravity}\label{Sec04:f(R)}

In recent years modifications of Einstein-Hilbert action by higher curvature terms is a subject of great interest. A very promising candidate among such modifications is the $f(\mathcal{R})$ gravity theory. The modifications introduced by the $f(\mathcal{R})$ term in the Lagrangian can address variety of problems. This theory also has the potential to survive all known tests of general relativity. It is therefore worthwhile to explore the nature of the effective gravitational field equation on a brane where the bulk is endowed with $f(\mathcal{R})$ gravity. 

\subsection{Effective Field Equation on the (n-1)-dimensional Brane}

We consider a $n$-dimension bulk spacetime endowed with $f(\mathcal{R})$ gravity. Following the standard procedure i.e., starting from the Gauss-Codazzi equation, subsequently using the $Z_{2}$ symmetry, we can obtain effective gravitational field equation on the brane. This has already been derived earlier in the context of a 4-dimensional brane embedded in a 5-dimensional bulk (see \cite{Chakraborty2015a} and the references therein). The effective gravitational field equation on the $(n-1)$-dimensional brane turns out to be:
\begin{align}
^{(n-1)}G_{\mu \nu}&=\kappa _{n-1}^{2}T^{(n-1)}_{\mu \nu}
\end{align}
where the stress-energy tensor $T^{(n-1)}_{\mu \nu}$ appearing on the right hand side can be decomposed into several small pieces. These include effective $(n-1)$-dimensional cosmological constant induced from the $n$-dimensional cosmological constant, brane energy momentum tensor and its higher order contribution. Also due to the presence of $f(\mathcal{R})$ gravity in the bulk there is an extra term and all these terms along with the non-local bulk Weyl tensor turns out to be:
\begin{align}
T^{(n-1)}_{\alpha \beta}=-\Lambda _{n-1} h_{\alpha \beta}+\tau _{\alpha \beta}
+\frac{\kappa _{n}^{4}}{\kappa _{n-1}^{2}}\pi _{\alpha \beta}-\frac{1}{\kappa _{n-1}^{2}}E_{\alpha \beta}
+\frac{1}{\kappa _{n-1}^{2}}Q_{\alpha \beta}
\end{align}
where the additional term due to $f(\mathcal{R})$ gravity is $Q_{\mu \nu}$. This additional term $Q_{\mu \nu}$ has the following expression:
\begin{align}
Q_{\mu \nu}=\left[\frac{1}{4}\frac{f(\mathcal{R})}{f'(\mathcal{R})}-
\frac{1}{4}\mathcal{R}-\frac{2}{3}\frac{\square f'(\mathcal{R})}{f'(\mathcal{R})}+\frac{2}{3}\frac{\nabla _{a}\nabla _{b}f'(\mathcal{R})}{f'(\mathcal{R})} \right]h_{\mu \nu}+\frac{2}{3}\frac{\nabla _{a}\nabla _{b}f'(\mathcal{R})}{f'(\mathcal{R})}e^{a}_{\mu}e^{b}_{\nu}
\end{align}
Note that in the \EH limit i.e. $f(\mathcal{R})\rightarrow \mathcal{R}$ the above term identically vanishes and the effective equation reduces to \eqs{GC11} and (\ref{GC12}). In the expression of the energy momentum tensor, $\Lambda _{n-1}$ is the $(n-1)$-dimensional brane cosmological constant (for explicit expression see \eq{GC13}), $\tau _{\mu \nu}$ is the brane energy momentum tensor, $\pi _{\alpha \beta}$ contain higher order terms like $\tau ^{\mu}_{\alpha}\tau _{\beta \mu}$ etc. (see \eq{GC10} for detailed expression). Finally $E_{\mu \nu}$ represents the non-local bulk effect. 

In order to simplify the expression for $Q_{\mu \nu}$, we can impose some assumptions. The simplification can be significant with the choice: $\partial _{\mu}\mathcal{R}=0$. Then using Taylor series expansion of bulk curvature $\mathcal{R}$ around $y=0$ hypersurface leads to: $\mathcal{R}=\mathcal{R}_{0}+\mathcal{R}_{1}y+\mathcal{R}_{2}y^{2}/2+\mathcal{O}(y^{3})$. As we have assumed that the bulk curvature depends only on the extra dimension $y$, all the coefficients appearing in the Taylor series expansion are constants. Hence we can conclude that all the derivatives calculated at $y=0$ yield a constant contribution which does not depend on any of the brane coordinates. In this case we can rewrite the stress-energy tensor as:
\begin{align}
T^{(n-1)}_{\alpha \beta}=-\Lambda _{\textrm{eff}} h_{\alpha \beta}+\tau _{\alpha \beta}
+\frac{\kappa _{n}^{4}}{\kappa _{n-1}^{2}}\pi _{\alpha \beta}-\frac{1}{\kappa _{n-1}^{2}}E_{\alpha \beta}
\end{align}
where we have an effective cosmological constant,
\begin{align}
\Lambda _{\textrm{eff}}=\Lambda _{n-1}-\frac{F(\mathcal{R})}{\kappa _{n-1}^{2}}
\end{align}
Thus the cosmological constant gets modified due to $f(\mathcal{R})$ gravity in the bulk leading to an additional constant term $F(\mathcal{R})$ to the induced cosmological constant $\Lambda _{n-1}$. This can provide a possible explanation to the cosmological constant problem by fine tuning $\Lambda _{n-1}$ with $F(\mathcal{R})$ term where the quantity $F(\mathcal{R})$ has the following expression:
\begin{align}\label{GCEqNew03}
F(\mathcal{R})=\left(g(\mathcal{R})+\frac{2}{3}\frac{\nabla _{A}\nabla _{B}f'(\mathcal{R})}
{f'(\mathcal{R})}n^{A}n^{B}\right)_{y=0}
\end{align}
Having derived the effective gravitational field equation in $(n-1)$-dimensional brane we will now write down the counterpart of this relation in $(n-m)$-dimensional brane as well.
\subsection{Generalization to (n-m)-dimensional Brane }

In this section we will generalize the above setup to an arbitrary lower dimensional hypersurface, namely, the $(n-m)$-dimensional brane. This can be done easily by taking a cue from the discussion in Einstein gravity. The effective field equation in the $(n-m)$-dimensional brane leads to the following expression:
\begin{align}\label{GCNew03}
^{(n-m)}G_{AB}=\kappa _{n-m}^{2}T_{AB}^{(n-m)}
\end{align}
with the following expression for the energy-momentum tensor as:
\begin{align}\label{GCNew04}
T_{AB}^{(n-m)}&=-\Lambda _{n-m}q_{AB}^{(n-m)}+\tau _{AB}^{(n-m)}+\frac{\kappa _{n-m+1}^{4}}{\kappa _{n-m}^{2}}\Upsilon _{AB}^{(n-m)}+\frac{1}{\kappa _{n-m}^{2}}Q_{AB}^{(n-m)}-\frac{1}{\kappa _{n-m}^{2}}E_{AB}^{(n-m)}
\nonumber
\\
&+\sum _{i=2}^{m}\left(\frac{n-i-2}{n-i-1}\right)\Big[\frac{\kappa _{n-i+1}^{2}}{\kappa _{n-i}^{2}} \left(t_{\alpha \beta}^{(n-i+1)}e^{\alpha}_{A}e^{\beta}_{B}
+q_{AB}\left\lbrace t_{\alpha \beta}^{(n-i+1)}s^{\alpha}s^{\beta}-\frac{1}{n-2}t^{(n-i+1)} \right\rbrace \right)
\nonumber
\\
&+\frac{\kappa _{n-i+2}^{4}}{\kappa _{n-i}^{2}}\left(\Pi _{\alpha \beta}^{(n-i+1)}e^{\alpha}_{A}e^{\beta}_{B}
+q_{AB}\left\lbrace \Pi _{\alpha \beta}^{(n-i+1)}s^{\alpha}s^{\beta}-\frac{1}{n-2}\Pi ^{(n-i+1)} \right\rbrace \right)
\nonumber
\\
&+\frac{1}{\kappa _{n-i}^{2}} \left(Q_{\alpha \beta}^{(n-i+1)}e^{\alpha}_{A}e^{\beta}_{B}
+q_{AB}\left\lbrace Q_{\alpha \beta}^{(n-i+1)}s^{\alpha}s^{\beta}-\frac{Q^{(n-i+1)}}{n-2} \right\rbrace \right)
\nonumber
\\
&-\frac{1}{\kappa _{n-i}^{2}} \left(E_{\alpha \beta}^{(n-i+1)}e^{\alpha}_{A}e^{\beta}_{B}
+q_{AB}\left\lbrace E_{\alpha \beta}^{(n-i+1)}s^{\alpha}s^{\beta}-\frac{E^{(n-i+1)}}{n-2} \right\rbrace \right)\Big]
\end{align}
Here $\Lambda _{n-m}$ is the induced brane cosmological constant, $\tau _{AB}^{(n-m)}$ represents the brane energy momentum tensor and $\Upsilon _{AB}^{(n-m)}$ is the higher order term. All the other extra terms originate from the effect of higher dimensional energy momentum tensor, $f(\mathcal{R})$ gravity and non-local effects from the bulk. However just as in the previous section in this case also we can simplify the expression further by assuming that bulk curvature depends only on the extra dimension $y$. Also if we assume that there is no energy momentum tensor on the higher dimensional branes and the bulk is AdS, then Weyl tensor vanishes identically. The effective gravitational field equation, then takes a particularly simple form. In this case following \cite{Chakraborty2015a}, we arrive at:
\begin{align}
^{(n-m)}G_{AB}=\kappa _{n-m}^{2}\left(-\Lambda _{n-m}q_{AB}^{(n-m)}+\tau _{AB}^{(n-m)}+\frac{\kappa _{n-m+1}^{4}}{\kappa _{n-m}^{2}}\Upsilon _{AB}^{(n-m)}\right)
\end{align}
where $\Lambda _{n-m}$ has to be constructed from $\Lambda _{\textrm{eff}}=\Lambda _{n-1}-F(\mathcal{R})$ by standard procedure illustrated for Einstein gravity. Thus in this case the effective equation would be identical to that obtained from general relativity, except for a modified cosmological constant. Also $\kappa _{n-m}^{2}$ is related to $(n-m)$-dimensional gravitational constant, $\tau _{AB}^{(n-m)}$ is the matter energy-momentum tensor on the $(n-m)$-dimensional brane, with $\Upsilon _{AB}^{(n-m)}$ being its higher order extension.

This completes our discussion on effective equation. However as an illustration we will consider a simple application of our work in a cosmological context in the next section.
\section{Applications to f(R) gravity}\label{Sec:Application}

Having discussed effective field equation in $f(\mathcal{R})$ gravity theory on a lower dimensional brane, alike general relativity we apply this in a static spherically symmetric and cosmological context. For that purpose we start with $6$-dimensional bulk spacetime and try to obtain a solutions from the effective field equations on a 4-dimensional brane. 
\subsection{Static Spherically Symmetric f(R) brane}

We start by considering the case for static, spherically symmetric spacetime with no matter source present. Thus brane energy momentum tensor vanishes however the electric part of the Weyl tensor survives. Hence the effective gravitational field equation on the four dimensional brane starting from the six dimensional bulk can be obtained from \eqs{GCNew03} and (\ref{GCNew04}) as,
\begin{align}
^{(4)}G_{AB}=-\frac{1}{2}\kappa _{5}^{2}\left(-\frac{1}{\kappa _{5}^{2}}F(\mathcal{R})+\frac{1}{6}\kappa _{5}^{2}\Sigma ^{2}\right)q_{AB}-\mathcal{E}_{AB}-\frac{2}{3}\left[E_{\alpha \beta}e^{\alpha}_{A}e^{\beta}_{B}+q_{AB}\left(E_{\alpha \beta}s^{\alpha}s^{\beta}-\frac{1}{4}E\right)\right]
\end{align}
where $F(\mathcal{R})$ is given in \eq{GCEqNew03}. In arriving at the above equation the cosmological constant has been neglected. The electric part of the Weyl tensor in such a static, spherically symmetric spacetime can be written explicitly using two unknown functions depending on the radial coordinate as,
\begin{align}
\mathcal{E}_{AB}=-\frac{\kappa _{5}^{4}}{\kappa _{4}^{4}}\left[U(r)\left(u_{A}u_{B}+\frac{1}{3}\bar{q}_{AB}\right)+P(r)\left(r_{A}r_{B}-\frac{1}{3}\bar{q}_{AB}\right)\right]
\end{align}
with $\bar{q}_{AB}$ being the induced metric on the $t=\textrm{constant}$ surface, such that $u^{A}\bar{q}_{AB}=0$. The five dimensional electric part of Weyl tensor would have similar decomposition depending on two separate radial functions $\bar{U}(r)$ and $\bar{P}(r)$. On projecting this five dimensional tensor $E_{\alpha \beta}$ on the four dimensional brane, the effective field equations reduce to
\begin{align}
G_{AB}&=-\frac{1}{2}\kappa _{5}^{2}\left(-\frac{1}{\kappa _{5}^{2}}F(\mathcal{R})+\frac{1}{6}\kappa _{5}^{2}\Sigma ^{2}\right)q_{AB}+\frac{\kappa _{5}^{4}}{\kappa _{4}^{4}}\left[U(r)\left(u_{A}u_{B}+\frac{1}{3}\bar{q}_{AB}\right)+P(r)\left(r_{A}r_{B}-\frac{1}{3}\bar{q}_{AB}\right)\right]
\nonumber
\\
&+\frac{2}{3}\frac{\kappa _{6}^{4}}{\kappa _{5}^{4}}\left[\bar{U}(r)\left(u_{A}u_{B}+\frac{1}{2}\bar{q}_{AB}\right)+\bar{P}(r)\left(r_{A}r_{B}-\frac{1}{2}\bar{q}_{AB}\right)\right]
\end{align}
Now $\kappa _{n}$ and $\kappa _{n-1}$ are related through \eq{GC14}, such that, $\kappa _{6}^{4}=(16/3)\kappa _{5}^{2}(1/\sigma)$, $\kappa _{5}^{4}=(6/\Sigma)\kappa _{4}^{2}$ and $\kappa _{4}^{2}=8\pi G$. Using these relations the above equation can be written in terms of the brane tensions, $\sigma$ for the five dimensional brane, $\Sigma$ for the four dimensional brane and the Gravitational constant $G$ as,
\begin{align}
G_{AB}&=-\left(-\frac{1}{2}F(\mathcal{R})+4\pi G \Sigma \right)q_{AB}+\frac{3}{4\pi G \Sigma}\left[U(r)\left(u_{A}u_{B}+\frac{1}{3}\bar{q}_{AB}\right)+P(r)\left(r_{A}r_{B}-\frac{1}{3}\bar{q}_{AB}\right)\right]
\nonumber
\\
&+\frac{32}{9\sqrt{6}}\frac{\sqrt{\Sigma}}{\sigma \sqrt{8\pi G}}\left[\bar{U}(r)\left(u_{A}u_{B}+\frac{1}{2}\bar{q}_{AB}\right)+\bar{P}(r)\left(r_{A}r_{B}-\frac{1}{2}\bar{q}_{AB}\right)\right]
\end{align}
Now assuming a static, spherically symmetric metric ansatz of the form,
\begin{align}
ds^{2}=-e^{\nu (r)}dt^{2}+e^{\lambda (r)}dr^{2}+r^{2}d\Omega ^{2}
\end{align}
we obtain various components of the gravitational field equations to be
\begin{align}
-e^{-\lambda}\left(-\frac{\lambda '}{r}+\frac{1}{r^{2}}\right)+\frac{1}{r^{2}}=\left(-\frac{1}{2}F(\mathcal{R})+4\pi G \Sigma \right)&+\frac{3}{4\pi G \Sigma}U(r)+\frac{32}{9\sqrt{6}}\frac{\sqrt{\Sigma}}{\sigma \sqrt{8\pi G}}\bar{U}(r)
\label{GCEqNew04}
\\
e^{-\lambda}\left(\frac{\nu '}{r}+\frac{1}{r^{2}}\right)-\frac{1}{r^{2}}=-\left(-\frac{1}{2}F(\mathcal{R})+4\pi G \Sigma \right)&+\frac{1}{4\pi G \Sigma}\left[U(r)+2P(r)\right]
\nonumber
\\
&+\frac{16}{9\sqrt{6}}\frac{\sqrt{\Sigma}}{\sigma \sqrt{8\pi G}}\left[\bar{U}(r)+\bar{P}(r)\right]
\label{GCEqNew05}
\\
\frac{1}{2}e^{-\lambda}\left(\nu ''+\frac{\nu '^{2}}{2}+\frac{\nu '-\lambda '}{r}-\frac{\nu '\lambda '}{2}\right)=-\left(-\frac{1}{2}F(\mathcal{R})+4\pi G \Sigma \right)&+\frac{1}{4\pi G \Sigma}\left[U(r)-P(r)\right]
\nonumber
\\
&+\frac{16}{9\sqrt{6}}\frac{\sqrt{\Sigma}}{\sigma \sqrt{8\pi G}}\left[\bar{U}(r)-\bar{P}(r)\right]
\end{align}
Hence the effect of multiple extra dimensions is manifested in the lower dimensional effective gravitational field equations through the appearance of extra terms $\bar{U}(r)$ and $\bar{P}(r)$. Also due to $f(\mathcal{R})$ gravity in the bulk there is one more additional term in the effective field equations, namely $F(\mathcal{R})$. In general, these equations has to be supplemented by the conservation equation, which will provide a differential relation connecting $U(r)$, $P(r)$, $\bar{U}(r)$ and $\bar{P}(r)$ which amounts to,
\begin{align}
\frac{1}{3}\left(\frac{dU}{dr}+2\frac{dP}{dr}\right)+\frac{\nu '}{3}\left(2U+P\right)+\frac{2P}{r}&=0
\label{GC_Conserv01a}
\\
\frac{1}{2}\left(\frac{d\bar{U}}{dr}+\frac{d\bar{P}}{dr}\right)+\frac{\nu '}{4}\left(3\bar{U}+\bar{P}\right)+\frac{2\bar{P}}{r}&=0
\label{GC_Conserv02a}
\end{align}
Alike the situation in \EH action all the objects on the right hand side of \eq{GCEqNew04} depend only on the radial coordinate which can be integrated leading to the following solution for $e^{-\lambda}$,
\begin{align}
e^{-\lambda}&=1-\frac{2GM}{r}-\frac{Q(r)}{r}-\left(-\frac{1}{2}F(\mathcal{R})+4\pi G \Sigma \right)\frac{r^{2}}{3}
\\
Q(r)&=\frac{3}{4\pi G \Sigma}\int dr ~r^{2}U(r)+\frac{32}{9\sqrt{6}}\frac{\sqrt{\Sigma}}{\sigma \sqrt{8\pi G}}\int dr ~r^{2}\bar{U}(r)
\end{align}
In order to obtain the solution for $e^{\nu}$ we need to know the solutions for the dark radiation and dark pressure terms. For that we will use the equations of states, $2U+P=0$ and $\bar{U}+\bar{P}=0$, such that,
\begin{align}
P(r)&=-2U(r)=\frac{P_{0}}{r^{4}};\qquad \bar{P}(r)=-3\bar{U}(r)=\frac{\bar{P}_{0}}{r^{6}}
\\
Q(r)&=Q_{0}+\frac{3\alpha P_{0}}{2r}+\frac{\bar{P}_{0}\beta}{9r^{3}};\qquad \alpha =\frac{1}{4\pi G\Sigma};\qquad \beta =\frac{32}{9\sqrt{6}}\frac{\sqrt{\Sigma}}{\sigma \sqrt{8\pi G}}
\end{align}
Then the full static spherically symmetric solution turns out to be,
\begin{align}
e^{\nu}=e^{-\lambda}=1-\frac{2GM+Q_{0}}{r}-\frac{3\alpha P_{0}}{2r^{2}}-\frac{\bar{P}_{0}\beta}{9r^{4}}-\left(-\frac{1}{2}F(\mathcal{R})+4\pi G \Sigma \right)\frac{r^{2}}{3}
\end{align}
This is an exact solution to the effective field equations where no assumptions or approximations have been made. The effect of extra spatial dimension appears through the charge term, i.e., $r^{-2}$ term and the $r^{-4}$ term and the effect of modified gravity appears through the $r^{2}$ term. Note that if we assume $F(\mathcal{R})=8\pi G\Sigma$, then our solution will coincide with the \EH action. Hence the addition of higher curvature terms to the bulk action can provide a potential solution to the cosmological constant problem.

The static spherically symmetric solution being dS or AdS has three important aspects to explore, namely --- (a) stability of the solutions, (b) anti-evaporation and (c) black hole thermodynamics. We will now comment on these features. The stability of solution can be ascertained by studying perturbations around it. These are again of three types, scalar, vector and tensor modes. These modes satisfy a set of decoupled wave equations with the following structure,
\begin{align}
\left(\square -\frac{1}{f(r)}V\right)\Phi =0
\end{align}
Here $\square$ stands for d'Alembertian operator, $f(r)=e^{\nu}=e^{-\lambda}$, $\Phi$ represents the perturbation modes and $V$ is an appropriate potential connected to the perturbation $\Phi$. For stability, we require the potentials to be positive. It turns out that for all choices of $F(\mathcal{R})$ the scalar and tensor modes have positive potentials. While for the vector mode we have stability for $F(\mathcal{R})>8\pi G\Sigma$, while the vector modes are unstable when $F(\mathcal{R})<8\pi G\Sigma$.

The anti-evaporation effect of this solution comes into picture when we set $P_{0}=0=\bar{P}_{0}$. Then the spherically symmetric solution represents a Schwarzschild (A)de-Sitter spacetime, which under proper limit can be mapped to the Nariai spacetime. A black hole in the Nariai spacetime has the peculiar property of increasing surface area due to quantum corrections as shown by Bousso and Hawking \cite{Bousso1998,Nojiri1999}. This phenomenon of anti-evaporation subsequently was generalized for Nariai black holes in the context of $f(\mathcal{R})$ gravity \cite{Nojiri2013}. Here $f(\mathcal{R})$ gravity plays the role of the anomaly induced effective action leading to anti-evaporation of black holes. In our case as well with $P_{0}=0=\bar{P}_{0}$, we have Nariai black hole as a limit. Thus from previous discussions it is clear that our solutions will also exhibit anti-evaporation effect. However for $P_{0}$ and $\bar{P}_{0}$ being not equal to zero, our solution cannot be reduced to the Nariai form and thus 
in general the solution presented here will not exhibit anti-evaporation.

Finally, the case of black hole thermodynamics is at our hand. The horizon associated with the above spacetime structure has the location $r=r_{h}$, which one can determine the following equation,
\begin{align}
1-\frac{2GM+Q_{0}}{r_{h}}-\frac{3\alpha P_{0}}{2r_{h}^{2}}-\frac{\bar{P}_{0}\beta}{9r_{h}^{4}}-\left(-\frac{1}{2}F(\mathcal{R})+4\pi G \Sigma \right)\frac{r_{h}^{2}}{3}=0
\end{align}
The mass of the black hole (or, equivalently the internal energy) can be obtained readily from the above equation as,
\begin{align}\label{Eq01new}
M(r_{h})=\frac{r_{h}}{2}-\frac{Q_{0}}{2}-\frac{3\alpha P_{0}}{4r_{h}}-\frac{\bar{P}_{0}\beta}{6r_{h}^{3}}-\left(-\frac{1}{2}F(\mathcal{R})+4\pi G \Sigma \right)\frac{r_{h}^{3}}{6}
\end{align}
The entropy can be obtained from holography, i.e., $S=k_{B}A/4\hbar$. In this particular case of spherically symmetric spacetime choosing $\hbar$ and Boltzmann constant appropriately we obtain, $S=r_{h}^{2}$. Thus internal energy of the system can be obtained in terms of the entropy from \eq{Eq01new} as,
\begin{align}
M(S)=\frac{\sqrt{S}}{2}-\frac{Q_{0}}{2}-\frac{3\alpha P_{0}}{4\sqrt{S}}-\frac{\bar{P}_{0}\beta}{6S^{3/2}}-\left(-\frac{1}{2}F(\mathcal{R})+4\pi G \Sigma \right)\frac{S^{3/2}}{6}
\end{align}
Then the black hole temperature can be obtained in terms of entropy as,
\begin{align}
T(S)=\left(\frac{\partial M}{\partial S}\right)=\frac{1}{4\sqrt{S}}+\frac{3\alpha P_{0}}{8S^{3/2}}+\frac{\bar{P}_{0}\beta}{4S^{5/2}}-\left(-\frac{1}{2}F(\mathcal{R})+4\pi G \Sigma \right)\frac{\sqrt{S}}{4}
\end{align}
which immediately leads to the following expression for specific heat as,
\begin{align}
C_{V}=T\left(\frac{\partial S}{\partial T}\right)=\frac{\frac{1}{4\sqrt{S}}+\frac{3\alpha P_{0}}{8S^{3/2}}+\frac{\bar{P}_{0}\beta}{4S^{5/2}}+\left(\frac{1}{2}F(\mathcal{R})-4\pi G \Sigma \right)\frac{\sqrt{S}}{4}}{\left(\frac{1}{2}F(\mathcal{R})-4\pi G \Sigma \right)\frac{1}{8\sqrt{S}}-\frac{1}{8S^{3/2}}-\frac{9\alpha P_{0}}{16S^{5/2}}-\frac{5\bar{P}_{0}\beta}{8S^{7/2}}}
\end{align}
It is clear from the above expression that the specific heat shows discontinuity in its behaviour, thereby indicating a second order phase transition. The surface of discontinuity in the specific heat is given by,
\begin{equation}\label{therm10}
\frac{1}{2}F(\mathcal{R})-4\pi G \Sigma=\frac{1}{S}+\frac{9\alpha P_{0}}{2S}+\frac{5\bar{P}_{0}\beta}{S^{5/2}}
\end{equation}
In order to understand physics behind this phase transition, it is always illuminating to discuss some limiting cases. For example, if we assume pure Einstein's gravity, the specific heat can be given by \eq{cvexp}, which is everywhere negative. This can also be divergent, provided the entropy becomes negative which cannot be satisfied in general. Thus bulk terms with positive dark pressure cannot lead to second order phase transition. In the other limit, where $P_{0}=0=\bar{P}_{0}$, we get the divergence of specific heat to correspond to the condition: $S^{-1}=(1/2)F(\mathcal{R})-4\pi G\Sigma$. Thus for $F(\mathcal{R})>8\pi G\Sigma$ we have second order phase transition. Our calculations therefore confirm that Schwarzschild Anti-de Sitter solution shows second order phase transition.

After deriving the static spherically symmetric solution along with its various properties, it is now time to discuss some novel features of this solution:
\begin{itemize}

\item The following restrictions provided by $2U(r)+P(r)=0$ and $\bar{P}(r)+3\bar{U}(r)=0$, shows that the metric elements satisfy the condition $e^{-\lambda}=e^{\nu}$. The line element so derived depends explicitly on the form of $U(r)$, $\bar{U}(r)$ and the gravity model $f(\mathcal{R})$. Thus the geometry of the spacetime is uniquely determined only when knowledge about multiple dimensions are obtained through $\bar{U}(r)$ and about the alternative gravity model. 

\item The solution is also stable under both scalar and tensor perturbations, since the potentials associated with respective wave equations are positive for scalar and tensor modes. Thus the black hole solution, will not change its character under scalar and tensor perturbations.

\item Finally, the specific heat associated with the black hole solution shows second order phase transition. This shows that the black hole transforms from thermodynamically unstable phase to thermodynamically stable phase. Moreover for vanishing of dark radiation we will obtain its mapping to Nariai black hole and therefore shows anti-evaporation.

\end{itemize}
Having discussed the effect on black hole solution, we will now discuss the cosmological solutions and its implication in detail. 
\subsection{Cosmological Scenario}

As in the case of general relativity, in cosmological scenario as well we can apply the effective field equations and obtain relevant solutions. Then we can check, where they are compatible with standard cosmological evolution. In the previous section we discussed vacuum solutions with Weyl tensor providing non-zero effect. In this we will consider exactly the opposite one, we will take $E_{ab}=0$ i.e. contribution from electric part of Weyl tensor is vanishing but there is matter in the spacetime. Then effective field equation in 4-dimension takes the following form:
\begin{align}
G_{AB}&=\left(-\frac{1}{2}F(\mathcal{R})+4\pi G \Sigma \right)q_{AB}+\kappa _{4}^{2}\tau _{AB}+\kappa _{5}^{4}\Upsilon _{AB}
\nonumber
\\
&+\frac{2}{3}\kappa _{5}^{2}\left(t_{\alpha \beta}e^{\alpha}_{A}e^{\beta}_{B}
+q_{AB}\left\lbrace t_{\alpha \beta}s^{\alpha}s^{\beta}-\frac{1}{4}t\right\rbrace \right)
\nonumber
\\
&+\frac{2}{3}\kappa _{6}^{4}\left(\Pi _{\alpha \beta}e^{\alpha}_{A}e^{\beta}_{B}
+q_{AB}\left\lbrace \Pi _{\alpha \beta}s^{\alpha}s^{\beta}-\frac{1}{4}\Pi \right\rbrace \right)
\end{align} 
In this expression along with brane energy momentum tensor $\tau _{AB}$ and its higher order term $\Upsilon _{AB}$, two additional contribution from 5-dimensional matter energy-momentum tensor is present. We assume that the brane is filled with perfect fluid with the following form for the energy-momentum tensor on the 4-dimension as:
\begin{align}
\tau ^{A}_{B}=\textrm{diag}\left(-\rho ,p,p,p\right)
\end{align}
where $\rho$ represents energy density of the matter fields and $p$ yields its pressure. The energy-momentum tensor in 5-dimension is taken to be pressure free dust, such that its energy momentum tensor has the simple form:
\begin{align}
t^{\alpha}_{\beta}=\textrm{diag}\left(-\rho _{5},0,0,0,0\right)
\end{align}
where the subscript `5' indicates that it is matter from 5-dimensional brane. Thus we arrive at the following expression for the higher order components:
\begin{align}
\Pi _{tt}=\frac{3}{32}\rho _{5}^{2};\qquad \Pi =-\frac{3}{32}\rho _{5}^{2}
\end{align}
Having obtained all the components in the above effective field equation, we can obtain the effective field equation itself. Both the time-time and spatial components of effective equation lead to:
\begin{align}
3H^{2}&=-\left(-\frac{1}{2}F(\mathcal{R})+4\pi G \Sigma \right)+\kappa _{4}^{2}\Lambda _{4}+\kappa _{4}^{2}\rho +\frac{1}{12}\kappa _{5}^{4}\rho ^{2}+\frac{1}{2}\kappa _{5}^{2}\rho _{5}+\frac{3}{64}\kappa _{6}^{4}\rho _{5}^{2}
\nonumber
\\
H^{2}+2\frac{\ddot{a}}{a}&=-\left(-\frac{1}{2}F(\mathcal{R})+4\pi G \Sigma \right)-\kappa _{4}^{2}p -\frac{1}{12}\kappa _{5}^{4}\left(\rho ^{2}+2p\rho \right)-\frac{1}{6}\kappa _{5}^{2}\rho _{5}-\frac{1}{64}\kappa _{6}^{4}\rho _{5}^{2}
\end{align}
For the bulk matter we have the following relation, namely, $\rho _{5}=(\rho _{5,0}/\ell)(1/a^{3})$, where $\ell$ represents the finite length of the 5-dimensional brane coordinate. After obtaining the Friedmann equations on the brane let us write them down in terms of relative abundances $\Omega _{i}=\rho _{i,0}/\rho _{c}$. The energy density appearing on the right hand side of Friedmann equations can be expanded in terms of baryonic and relativistic matter fields. For higher dimensional matter being non-relativistic the Hubble parameter turns out to be,
\begin{align}
H^{2}=H_{0}^{2}\left[\Omega _{\Lambda}+\left(\Omega _{B}+\Omega _{5}\right)\left(\frac{a_{0}}{a}\right)^{3}+\Omega _{R}\left(\frac{a_{0}}{a}\right)^{4}\right]
\end{align}
where, $\Omega _{5}=(\kappa _{5}^{2}/16\pi G \ell)(\rho _{5,0}/\rho _{c})$ and $\Omega _{\Lambda}$ is from the combination $[-(1/2)F(\mathcal{R})+4\pi G \Sigma]$. Thus we have reproduced the standard $\Lambda CDM$ model. At late times $\Omega _{\Lambda}$ would dominate leading to accelerated expansion of the universe, however this acceleration is originating since the bulk spacetime is being described by $f(\mathcal{R})$ gravity theory and we are living on a brane with nonzero brane tension (the $\Sigma$ term). Starting from the $1/a^{4}$ term, leading to radiation dominated universe, we have matter (i.e., non-relativistic matter) dominated epoch such that $H^{2}\sim a^{-3}$ and finally the accelerating sector. The intriguing fact about the matter sector is that along with the baryonic matter, due to existence of extra spatial dimensions we have an additional contribution. This additional contribution might act as a dark matter candidate. In standard $\Lambda CDM$ cosmology, the dark matter has to be 
included by hand while in our model it appears \emph{naturally}. To get an estimate of five dimensional parameters we can use the epoch of matter, radiation equality leading to,
\begin{align}
1+z_{eq}=3.9\times 10^{4}\left(0.022+\frac{\kappa _{5}^{2}h^{2}}{16\pi G \ell}\frac{\rho _{5,0}}{\rho _{c}}\right)
\end{align}
where $h=0.72$, $\ell$ stands for the length of the extra dimension. From the redshift $z_{eq}$ one can determine the value of the higher dimensional gravitational constant $\kappa _{5}^{2}$. On the other hand, unlike Einstein-Hilbert bulk in this case as well by tuning the $f(\mathcal{R})$ model itself it is possible to have inflation. Thus the effective field equations on the brane are capable to explain the various phases of the universe.

Let us now explore the effect of additional correction terms on the Hubble parameter. To confine ourselves to one representative example, if we assume the matter in the 4-dimensional brane to be pressure-less then we have,
\begin{align}
H^{2}=-\frac{1}{3}\left(-\frac{1}{2}F(\mathcal{R})+4\pi G \Sigma \right)+\frac{8\pi G}{3}\frac{\rho _{0}}{a^{3}} +\frac{1}{36}\kappa _{5}^{4}\frac{\rho _{0}^{2}}{a^{6}}+\frac{1}{6}\kappa _{5}^{2}\frac{\rho _{5,0}}{\ell}\frac{1}{a^{3}}+\frac{1}{64}\kappa _{6}^{4}\frac{\rho _{5,0}^{2}}{\ell ^{2}}\frac{1}{a^{6}}
\end{align}
where $\rho _{0}$ represents the value of the energy density in the present epoch and $\rho _{5,0}$ represents the same for 5-dimensional matter. The above expression for Hubble parameter can be simplified and arranged properly leading to:
\begin{align}\label{HubbleMultiple}
H^{2}=-\frac{1}{3}\left(-\frac{1}{2}F(\mathcal{R})+4\pi G \Sigma \right)+\left\lbrace \frac{8\pi G}{3}\rho _{0}+\frac{1}{6}\kappa _{5}^{2}\frac{\rho _{5,0}}{\ell} \right\rbrace \frac{1}{a^{3}}+\left\lbrace \frac{1}{36}\kappa _{5}^{4}\rho _{0}^{2}+\frac{1}{64}\kappa _{6}^{4}\frac{\rho _{5,0}^{2}}{\ell ^{2}}\right\rbrace \frac{1}{a^{6}}
\end{align}
Following the techniques developed in \cite{Chakraborty2014d} we can introduce three constants, namely:
\begin{align}
C_{f}=-\frac{1}{3}\left(-\frac{1}{2}F(\mathcal{R})+4\pi G \Sigma \right);\qquad C_{\rho}=\frac{8\pi G}{3}\rho _{0}+\frac{1}{6}\kappa _{5}^{2}\frac{\rho _{5,0}}{\ell};\qquad C_{\textrm{Sq}}=\frac{1}{36}\kappa _{5}^{4}\rho _{0}^{2}+\frac{1}{64}\kappa _{6}^{4}\frac{\rho _{5,0}^{2}}{\ell ^{2}}
\end{align}
such that the following solution to the scale factor is obtained,
\begin{align}\label{GC72}
\exp \left[3\sqrt{C_{f}}\left(t-t_{0}\right)\right]&=
\frac{2\sqrt{C_{f}}\sqrt{C_{f}a^{6}+C_{\rho}a^{3}+C_{\textrm{Sq}}}+2C_{f}a^{3}+C_{\rho}}
{2\sqrt{C_{f}}\sqrt{C_{f}+C_{\rho}+C_{\textrm{Sq}}}+2C_{f}+C_{\rho}};\qquad C_{f}>0
\\
3\sqrt{-C_{f}}\left(t_{0}-t\right)&=
\sin ^{-1} \left(\frac{2C_{f}a^{3}+C_{\rho}}{\sqrt{C_{\rho}^{2}-4C_{f}C_{\textrm{Sq}}}}\right)
-\sin ^{-1} \left(\frac{2C_{f}+C_{\rho}}{\sqrt{C_{\rho}^{2}-4C_{f}C_{\textrm{Sq}}}} \right);
\nonumber
\\
\qquad C_{f}&<0,\qquad C_{\rho}^{2}>4C_{f}C_{\textrm{Sq}}
\end{align}
Though the solutions are a bit complicated, we can impose certain conditions on them such that these solutions simplify substantially. For that purpose we consider the situation, where $C_{f}=0$, i.e., $F(\mathcal{R})=8\pi G \Sigma$. Since $\Sigma$ is positive definite the above result suggests that $F(\mathcal{R})$ is positive. Applying this for matter dominated epoch, the Hubble parameter from \eq{HubbleMultiple} turns out to be,
\begin{equation}
H^{2}=\frac{8\pi G}{3}\rho \left[1+\frac{\kappa _{5}^{2}}{16\pi G}\frac{\rho _{5}}{\rho}-\frac{4\pi G\rho}{F(\mathcal{R})}
\left(1+\frac{9}{16}\frac{\kappa _{6}^{4}}{\kappa _{5}^{4}}\frac{\rho _{5}^{2}}{\rho ^{2}}\right)\right]
\end{equation}  
Assume $\rho$ to be energy density of non-relativistic matter we arrive at,
\begin{equation}
H^{2}=\left\lbrace \frac{8\pi G \rho _{0}}{3}+\frac{\kappa _{5}^{2}}{6}\frac{\rho _{5,0}}{\ell}\right\rbrace \frac{1}{a^{3}}+\frac{4\pi G \rho _{0}^{2}}{3\Sigma}
\left(1+\frac{9}{16}\frac{\kappa _{6}^{4}}{\kappa _{5}^{4}}\frac{\rho _{5,0}^{2}}{\ell ^{2}\rho _{0} ^{2}}\right)\frac{1}{a^{6}}
\end{equation}
leading to the following solution for the scale factor:
\begin{equation}\label{New03}
a^{3}=\left\lbrace 6\pi G \rho _{0}+3\frac{\kappa _{5}^{2}}{8}\frac{\rho _{5,0}}{\ell}\right\rbrace t^{2}+\left\lbrace \sqrt{\frac{12\pi G \rho _{0}^{2}}{\Sigma}
\left(1+\frac{9}{16}\frac{\kappa _{6}^{4}}{\kappa _{5}^{4}}\frac{\rho _{5,0}^{2}}{\ell ^{2}\rho _{0} ^{2}}\right)}\right\rbrace t
\end{equation}
It is clear from the above expression that the universe undergoes a transition in the expansion rate. The time scale when this happened depends on whether the brane matter energy dominates over the bulk matter or not. Below we present the result for timescale of transition for both the situations:  
\begin{align}
t\sim \sqrt{\frac{1}{3\pi G \Sigma}}&=\sqrt{-\frac{8}{3F(\mathcal{R})}};
\qquad (\rho _{5,0}/\ell \ll \rho _{0})
\\
t \sim & \frac{\kappa _{6}^{2}}{\kappa _{5}^{2}}\left(1-\frac{16\pi G\rho _{0}\ell}{\kappa _{5}^{2}\rho _{5,0}}\right)
\qquad (\rho _{5,0}/\ell \gg \rho _{0})
\end{align}
Thus at early universe we have a high energy regime, where $a\sim t^{1/3}$, while at late time low energy regime the scale factor variation with time modifies to $a\sim t^{2/3}$, which is the standard evolution of the matter field. Note that in Einstein gravity as well e had similar transition in the scale factor. But in this particular case, importantly the transition time depends crucially on the $f(\mathcal{R})$ model we are considering. Similar phenomenon will be observed for radiation dominated epoch as well.

The key aspects of the above results are summarized as follows:
\begin{itemize}

\item First and foremost, the issue of dark energy in this particular case is resolved by the addition of higher curvature terms to the Einstein-Hilbert action and through the brane tension. The same $f(\mathcal{R})$ model can also explain the inflationary scenario without invoking any scalar field in the picture. 

\item The cosmology as presented by modified Friedmann equations automatically lead to an additional matter component originating from higher dimensions, which behave as a dark matter candidate. 

\item Finally, note that due to presence of additional correction terms in the effective equations on the brane few new cosmological phases enter the picture. For example, the matter dominated epoch gets divided into two parts and this depends explicitly on the $f(\mathcal{R})$ model under consideration. This modifications will have indirect effects on the structure formation, which can be used to constrain not only higher dimensional parameters but also the form of alternative gravity theories.
 
\end{itemize}

Thus with matter present in the 5-dimensional brane we can have standard cosmology with proper scaling of the scale factor with time. Moreover from the effective field equation it is clear that the 5-dimensional matter adds to the 4-dimensional one and enhances the total non-relativistic matter content of the universe (4-dimensional brane we live in). Structure of angular power spectrum, perturbation in this multiple extra dimension scenario can lead to interesting and important results by providing constraints on various parameters in this models. 

It is possible to ascribe the cosmological solutions so obtained holographic interpretations. Following  \cite{Verlinde2000} it suffices to show that the usual four dimensional cosmological equations can be re-written as an two dimensional conformal field theory (CFT). For this we can define the Hawking, Bekenstein and Bekenstein-Hawking entropies as,
\begin{align}
S_{H}=\frac{(n-1)HV}{4G},\qquad S_{BH}=\frac{(n-1)V}{4Ga},\qquad S_{B}=\frac{2\pi a}{n}E
\end{align}
where $E=\rho V$ is the total energy within a volume V and $\rho _{m}\propto a^{-n}$, but $\rho =\rho _{m} +F(\mathcal{R})/16\pi G$. This is how the effect of $f(\mathcal{R})$ gravity enters the picture, by altering the definition for energy. Furthermore the total entropy is by the Hubble entropy $S_{H}$, which is the entropy of the black hole with the radius of the Hubble size. Using these definitions the Friedman equations can be rewritten as,
\begin{align}
S_{H}=\frac{2\pi a}{n}\sqrt{E_{BH}\left(2E-KE_{BH}\right)};\qquad KE_{BH}=n\left(E+pV-T_{H}S_{H}\right)
\end{align}
with energy of the black hole being $E_{BH}=n(n-1)V/8\pi Ga^{2}$ and Hawking temperature being $T_{H}=-\dot{H}/2\pi H$. This is exactly the form one would have from the CV entropy relation. It explicitly shows the connection between four dimensional classical gravitational physics and 2-dimensional quantum 2-dimensional CFT \cite{Cardy1986}. One can also consider dS or AdS black holes in the bulk, such that the brane equations of motion can be casted in the form of Friedman equations. By introducing brane matter the cosmological equations can be written in the CV form which is related to AdS black hole entropy \cite{Nojiri2002}.
\section{Discussion}

Motivated by the success of extra dimensional models to explain various physical problems, e.g., hierarchy problem, cosmological constant problem, in this work we have constructed and worked within the premises of a very general higher dimensional model. Since gravity can propagate in the full bulk spacetime, the field equations for gravity on a lower dimensional brane is taken to be induced from the higher dimensional one. Previous works in this direction involve two assumptions, namely, the bulk is five-dimensional while the brane is four-dimensional and gravity in the bulk is originating from Einstein-Hilbert action. However in the spirit of string theory it is pertinent to ask what happens to this induced gravity program as the bulk spacetime is $n$-dimensional while the brane is $(n-m)$-dimensional, for arbitrary $m$. Also from the recent success of $f(\mathcal{R})$ gravity it is interesting to discuss the situation when the bulk spacetime has $f(\mathcal{R})$ as the action not $\mathcal{R}$. Below we 
summarize all the results obtained in this work along these lines:
\begin{itemize}
\item Starting from a n-dimensional bulk spacetime endowed with Einstein gravity, we first derive the effective field equations on $(n-1)$ and $(n-2)$ spacetime dimensions. Taking a cue from this analysis we could read off the effective field equations in $(n-m)$-dimensions for arbitrary $m$. 
\item As an application of this result, we have illustrated two situations. One, in which there is no matter but bulk effects populate the right hand side of effective field equations. The static spherically symmetric solution so obtained has the standard $1-(2GM/r)$ term, with additional terms originating from the bulk Weyl tensor. These additional terms have contributions from both the five-dimensional and six-dimensional Weyl tensors (in particular its electric part). Secondly, cosmological situation with matter but no bulk geometric effect. From the solutions so obtained it becomes evident that matter in higher dimensions can add to the matter content of the visible brane and has the potential to act as dark matter source. More importantly, the effective Friedman equations can be casted in a holographic language, such that it can be related to two dimensional CFT. 
\item Modifying Einstein-Hilbert action in the bulk by addition of $f(\mathcal{R})$ term is the next immediate generalization. In this context as well, we have first derived the effective field equations in $(n-1)$-dimensional brane starting from $n$-dimensional bulk. Then we have obtained the effective field equations on a $(n-m)$-dimensional brane, starting from the original $n$-dimensional bulk. 
\item Alike the situation in general relativity, in this context as well we have discussed two cases. For the spherically symmetric, static vacuum spacetime, $f(\mathcal{R})$ gravity modifies the general relativity solution significantly. The solution also exhibits two remarkable effects. Firstly, it shows divergent specific heat, which points out the effect of phase transition. Secondly, it can be written in Nariai form, which shows the anti-evaporation effect. 

\item In the cosmological context as well $f(\mathcal{R})$ gravity modifies the cosmological constant and the transition time scale depends explicitly on the $f(\mathcal{R})$ gravity model under consideration. Also the $f(\mathcal{R})$ gravity model exhibits the CV entropy relation, such that Friedman equations can be rewritten in terms of two dimensional CFT.
\end{itemize}

Thus in a nutshell, starting from Einstein's theory in the bulk, which is taken to be $n$-dimensional we have constructed various induced objects on any arbitrary $(n-m)$-dimensional brane providing a complete generalization of existing set up. We generalize this analysis further by incorporating $f(\mathcal{R})$ gravity in the bulk instead of Einstein's gravity. In $f(\mathcal{R})$ gravity as well we follow the same procedure by relating geometrical objects like extrinsic curvature, Riemann curvature in the $n$-dimensional bulk with the $(n-m)$-dimensional brane. In this case as well from the structure of the effective equation we can immediately generalize the set up on an arbitrary lower dimensional hypersurface as well. 

Having derived all the necessary theoretical ingredients, we consider possible applications of our results in both the gravity theories. As an illustration, we took a 6-dimensional bulk spacetime in which a 4-dimensional brane is embedded. In static spherically symmetric vacuum spacetime we have additional terms from bulk Weyl tensor as the source of gravity. For $f(\mathcal{R})$ gravity we have an equivalent of cosmological constant making the solutions asymptotically (Anti) de-Sitter.  In cosmological context the 5-dimensional sub-manifold contains matter showing that the standard cosmology to be an inherent and natural consequence of our model. Finally we comment on the possibility of the bulk matter as a possible candidate of dark matter in the lower dimensional brane.

\section*{Acknowledgement}

Research of S.C. is funded by a SPM fellowship from CSIR, Govt. of India.

\appendix

\section{Appendix: Detailed Calculations}

In this section, we provide some detailed expressions, which have been used in the main text for arriving at various results. 

\subsection{Identities for the Derivation of (n-1)-dimensional Effective Field Equation}\label{App1}

The starting point of obtaining the effective field equation is the Gauss-Codazzi equation, which relates the $(n-1)$-dimensional curvature tensor to $n$-dimensional curvature tensor. For our purpose these two tensors are related through the following relation:
\begin{align}\label{GCEq01}
^{(n-1)}R_{\alpha \beta \mu \nu}=~^{(n)}R_{abcd}e^{a}_{\alpha}e^{b}_{\beta}e^{c}_{\mu}e^{d}_{\nu}
-\epsilon \left(K_{\alpha \nu}K_{\beta \mu}-K_{\alpha \mu}K_{\beta \nu}\right)
\end{align}
where we have used the definition: 
\begin{equation}\label{GCEq02}
K_{\alpha \beta}=e^{a}_{\alpha}e^{b}_{\beta}\nabla _{a}n_{b}
\end{equation}
with $\epsilon =n_{i}n^{i}$. Contraction of \eq{GCEq01} with the $(n-1)$ dimensional metric $h_{\alpha \beta}=e^{a}_{\alpha}e^{b}_{\beta}\left(g_{ab}-\epsilon n_{a}n_{b}\right)$ leads to the connection between $(n-1)$ dimensional Ricci tensor to $n$ dimensional curvature tensor as,
\begin{align}\label{GCEq03}
^{(n-1)}R_{\alpha \mu}&=h^{\beta \nu}~^{(n-1)}R_{\alpha \beta \mu \nu}
\nonumber
\\
&=~^{(n)}R_{abcd}e^{a}_{\alpha}e^{b}_{\beta}e^{c}_{\mu}e^{d}_{\nu}h^{\beta \nu}
-\epsilon \left(K_{\alpha \nu}K_{\beta \mu}-K_{\alpha \mu}K_{\beta \nu}\right)h^{\beta \nu}
\nonumber
\\
&=~^{(n)}R_{abcd}e^{a}_{\alpha}e^{c}_{\mu}\left(g^{bd}-\epsilon n^{b}n^{d}\right)
-\epsilon \left(K_{\alpha \nu}K^{\nu}_{\mu}-K_{\alpha \mu}K\right)
\nonumber
\\
&=~^{(n)}R_{ac}e^{a}_{\alpha}e^{c}_{\mu}-\epsilon ~^{(n)}R_{abcd}e^{a}_{\alpha}n^{b}e^{c}_{\mu}n^{d}
-\epsilon \left(K_{\alpha \nu}K^{\nu}_{\mu}-K_{\alpha \mu}K\right)
\end{align}
The Ricci scalar can now be obtained as,
\begin{align}\label{GCEq04}
^{(n-1)}R&=h^{\alpha \mu}~^{(n-1)}R_{\alpha \mu}
\nonumber
\\
&=~^{(n)}R_{ac}e^{a}_{\alpha}e^{c}_{\mu}h^{\alpha \mu}
-\epsilon ~^{(n)}R_{abcd}e^{a}_{\alpha}n^{b}e^{c}_{\mu}n^{d}h^{\alpha \mu}
-\epsilon \left(K_{\alpha \nu}K^{\nu}_{\mu}-K_{\alpha \mu}K\right)h^{\alpha \mu}
\nonumber
\\
&=~^{(n)}R_{ac}\left(g^{ac}-\epsilon n^{a}n^{c}\right)
-\epsilon ~^{(n)}R_{abcd}n^{b}n^{d}\left(g^{ac}-\epsilon n^{a}n^{c}\right)
-\epsilon \left(K_{\mu \nu}K^{\mu \nu}-K^{2}\right)
\nonumber
\\
&=^{(n)}R-2\epsilon ~^{(n)}R_{ac}n^{a}n^{c}
-\epsilon \left(K_{\mu \nu}K^{\mu \nu}-K^{2}\right)
\end{align}
The above relation can be further simplified by using an identity for $^{(n)}R_{ab}n^{a}n^{b}$, which can be written as:
\begin{align}\label{GCEq05}
^{(n)}R_{ab}n^{a}n^{b}&=\nabla _{a}\left(n^{b}\nabla _{b}n^{a}\right)-\nabla _{a}n^{b}\nabla _{b}n^{a} -\nabla _{b}\left(n^{b}\nabla _{a}n^{a}\right)+\left(\nabla _{b}n^{b}\right)^{2}
\nonumber
\\
&=-\nabla _{i}\left(Kn^{i}-a^{i}\right)-K_{\alpha \beta}K^{\alpha \beta}+K^{2}
\end{align}
Then we obtain $^{(n)}R$ in terms of $^{(n-1)}R$ as:
\begin{align}\label{GCEq06}
^{(n)}R=^{(n-1)}R-\epsilon \left(K_{\mu \nu}K^{\mu \nu}-K^{2}\right)-2\epsilon \nabla _{i}\left(Kn^{i}-a^{i}\right)
\end{align}
The relation between $(n-1)$ dimensional Einstein tensor and $n$ dimensional Riemann tensor is therefore given by
\begin{align}\label{GCEq07}
^{(n-1)}G_{\alpha \mu}&=~^{(n-1)}R_{\alpha \mu}-\frac{1}{2}h_{\alpha \mu}~^{(n-1)}R
\nonumber
\\
&=~^{(n)}R_{ac}e^{a}_{\alpha}e^{c}_{\mu}-\epsilon ~^{(n)}R_{abcd}e^{a}_{\alpha}n^{b}e^{c}_{\mu}n^{d}
-\epsilon \left(K_{\alpha \nu}K^{\nu}_{\mu}-K_{\alpha \mu}K\right)
\nonumber
\\
&-\frac{1}{2}h_{\alpha \mu}\left[~^{(n)}R-2\epsilon ~^{(n)}R_{ac}n^{a}n^{c}
-\epsilon \left(K_{\mu \nu}K^{\mu \nu}-K^{2}\right)\right]
\nonumber
\\
&=\left(~^{(n)}R_{ac}-\frac{1}{2}g_{ac}~^{(n)}R\right)e^{a}_{\alpha}e^{c}_{\mu}
+\epsilon ~^{(n)}R_{ac}n^{a}n^{c}h_{\alpha \mu}
-\epsilon ~^{(n)}R_{abcd}e^{a}_{\alpha}n^{b}e^{c}_{\mu}n^{d}
\nonumber
\\
&+\epsilon \left[ KK_{\alpha \mu}-K_{\alpha \beta}K^{\beta}_{\mu}
-\frac{1}{2}h_{\alpha \mu}\left(K^{2}-K_{\mu \nu}K^{\mu \nu} \right)\right]
\end{align}
The Riemann tensor can be decomposed into Ricci tensor, Ricci scalar and Weyl curvature tensor in $n$ dimension as,
\begin{align}\label{GCEq08}
^{(n)}R_{abcd}&=~^{(n)}C_{abcd}+\frac{1}{n-2}\left(g_{ac}R_{bd}-g_{ad}R_{bc}
-g_{bc}R_{ad}+g_{bd}R_{ac}\right)
\nonumber
\\
&-\frac{1}{(n-1)(n-2)}\left(g_{ac}g_{bd}-g_{ad}g_{bc}\right)R
\end{align}
which suggests,
\begin{align}\label{GCEq09}
^{(n)}R_{abcd}e^{a}_{\alpha}n^{b}e^{c}_{\mu}n^{d}&=~^{(n)}C_{abcd}e^{a}_{\alpha}n^{b}e^{c}_{\mu}n^{d}
+\frac{1}{n-2}\left(R_{ac}n^{a}n^{c}h_{\alpha \mu}+\epsilon R_{ac}e^{a}_{\alpha}e^{c}_{\mu}\right)
\nonumber
\\
&-\frac{\epsilon}{(n-1)(n-2)}h_{\alpha \mu}R
\end{align}
Then substitution of the above equation in \eq{GCEq07} leads to,
\begin{align}\label{GCEq10}
^{(n-1)}G_{\alpha \beta}&=~^{(n)}G_{ab}e^{a}_{\alpha}e^{b}_{\beta}
+\epsilon \left(\frac{n-3}{n-2}\right)~^{(n)}R_{ab}n^{a}n^{b}h_{\alpha \beta}
-\frac{1}{n-2}~^{(n)}R_{ab}e^{a}_{\alpha}e^{b}_{\beta}
\nonumber
\\
&+\frac{1}{(n-1)(n-2)}~^{(n)}R~h_{\alpha \beta}-\epsilon ~^{(n)}E_{\alpha \beta}
\nonumber
\\
&+\epsilon \left[ KK_{\alpha \beta}-K_{\alpha \mu}K^{\mu}_{\beta}
-\frac{1}{2}h_{\alpha \beta}\left(K^{2}-K_{\mu \nu}K^{\mu \nu} \right)\right]
\end{align}
where we have defined, $^{(n)}E_{\alpha \beta}=^{(n)}C_{abcd}e^{a}_{\alpha}n^{b}e^{c}_{\beta}n^{d}$. This yields the $n$ dimensional Einstein equation, with the following expressions:
\begin{align}\label{GCEq11}
^{(n)}R_{ab}-\frac{1}{2}g_{ab}~^{(n)}R&=\kappa _{n}^{2}T_{ab};
\qquad
^{(n)}R=-\frac{2}{n-2}\kappa _{n}^{2}T;
\qquad
^{(n)}R_{ab}=\kappa _{n}^{2}\left(T_{ab}-\frac{1}{n-2}g_{ab}T\right)
\end{align}
We therefore have the following expression
\begin{align}\label{GCEq12}
^{(n)}G_{ab}e^{a}_{\alpha}e^{b}_{\beta}
&+\epsilon \left(\frac{n-3}{n-2}\right)~^{(n)}R_{ab}n^{a}n^{b}h_{\alpha \beta}
-\frac{1}{n-2}~^{(n)}R_{ab}e^{a}_{\alpha}e^{b}_{\beta}
+\frac{1}{(n-1)(n-2)}~^{(n)}R~h_{\alpha \beta}
\nonumber
\\
&=\kappa _{n}^{2}T_{ab}e^{a}_{\alpha}e^{b}_{\beta}
+\epsilon \kappa _{n}^{2}\left(\frac{n-3}{n-2}\right)
\left(T_{ab}-\frac{1}{n-2}g_{ab}T\right)n^{a}n^{b}h_{\alpha \beta}
\nonumber
\\
&-\kappa _{n}^{2}\frac{1}{n-2}\left(T_{ab}-\frac{1}{n-2}g_{ab}T\right)e^{a}_{\alpha}e^{b}_{\beta}
-\kappa _{n}^{2}\frac{2}{(n-1)(n-2)^{2}}h_{\alpha \beta}T
\nonumber
\\
&=\kappa _{n}^{2}\frac{n-3}{n-2}\left[T_{ab}e^{a}_{\alpha}e^{b}_{\beta}
+h_{\alpha \beta}\left\lbrace \epsilon T_{ab}n^{a}n^{b}-\frac{1}{n-1}T \right\rbrace \right]
\end{align}
This is the result used in \sect{Sec01:n-1}.

\subsection{Identities for the Derivation of (n-2)-dimensional Effective Field Equation}\label{App2}

We start with the connection between $(n-2)$ dimensional curvature tensor and $(n-1)$ dimensional curvature tensor as,
\begin{align}\label{GCEq13}
^{(n-2)}R_{ABCD}=~^{(n-1)}R_{\alpha \beta \mu \nu}e^{\alpha}_{A}e^{\beta}_{B}e^{\mu}_{C}e^{\nu}_{D}
-\left(\mathcal{K}_{AD}\mathcal{K}_{BC}-\mathcal{K}_{AC}\mathcal{K}_{BD} \right)
\end{align}
where we have the extrinsic curvature on the $(n-2)$ dimensional surface as: $\mathcal{K}_{AB}=e^{\alpha}_{A}e^{\beta}_{B}\nabla _{\alpha}s_{\beta}$, with $s_{\alpha}$ being the normal to the surface. Then the induced $(n-2)$ dimensional metric turns out to be, $q_{AB}=e^{\alpha}_{A}e^{\beta}_{B}(h_{\alpha \beta}-s_{\alpha}s_{\beta})$. Contracting the above equation with $q^{BD}$ leads to,
\begin{align}\label{GCEq14}
^{(n-2)}R_{AC}&=~^{(n-1)}R_{\alpha \beta \mu \nu}e^{\alpha}_{A}e^{\beta}_{B}e^{\mu}_{C}e^{\nu}_{D}q^{BD}
-\left(\mathcal{K}_{AD}\mathcal{K}^{D}_{C}-\mathcal{K}_{AC}\mathcal{K}\right)
\nonumber
\\
&=~^{(n-1)}R_{\alpha \beta \mu \nu}e^{\alpha}_{A}e^{\mu}_{C}\left(h^{\beta \nu}-s^{\beta}s^{\nu}\right)
-\left(\mathcal{K}_{AD}\mathcal{K}^{D}_{C}-\mathcal{K}_{AC}\mathcal{K}\right)
\nonumber
\\
&=~^{(n-1)}R_{\alpha \mu}e^{\alpha}_{A}e^{\mu}_{C}-
~^{(n-1)}R_{\alpha \beta \mu \nu}e^{\alpha}_{A}s^{\beta}e^{\mu}_{C}s^{\nu}
-\left(\mathcal{K}_{AD}\mathcal{K}^{D}_{C}-\mathcal{K}_{AC}\mathcal{K}\right)
\end{align}
On further contraction we arrive at the following result,
\begin{align}\label{GCEq15}
^{(n-2)}R&=q^{AB}~^{(n-2)}R_{AB}
\nonumber
\\
&=~^{(n-1)}R-2~^{(n-1)}R_{\alpha \mu}s^{\alpha}s^{\mu}-
\left(\mathcal{K}_{AB}\mathcal{K}^{AB}-\mathcal{K}^{2} \right)
\end{align}
Then we can use \eq{GCEq05} to obtain an identical relation for $^{(n-1)}R_{\alpha \mu}s^{\alpha}s^{\mu}$. This leads to the relation:
\begin{align}\label{GCEq16}
^{(n-1)}R=^{(n-2)}R-\left(\mathcal{K}_{AB}\mathcal{K}^{AB}-\mathcal{K}^{2}\right) -2D_{\alpha}\left(\mathcal{K}s^{\alpha}-s^{\beta}D_{\beta}s^{\alpha}\right)
\end{align}
which on using \eq{GCEq06} yields:
\begin{align}\label{GCEq17}
^{(n)}R&=^{(n-2)}R-\left(\mathcal{K}_{AB}\mathcal{K}^{AB}-\mathcal{K}^{2}\right) -2D_{\alpha}\left(\mathcal{K}s^{\alpha}-s^{\beta}D_{\beta}s^{\alpha}\right)
\nonumber
\\
&+\left(K_{\mu \nu}K^{\mu \nu}-K^{2}\right)+2\nabla _{i}\left(Kn^{i}-a^{i}\right)
\end{align}
In order to obtain the effective equation on this $(n-2)$ dimensional surface we need various contractions of the Riemann tensor. For that purpose we start with the following identity,
\begin{align}\label{GCEq18}
^{(n-1)}R_{\alpha \beta \mu \nu}e^{\alpha}_{A}s^{\beta}e^{\mu}_{B}s^{\nu}
&=\Big\lbrace ^{(n-1)}C_{\alpha \beta \mu \nu}
+\frac{1}{n-3}\left(h_{\alpha \mu}~^{(n-1)}R_{\beta \nu}-h_{\alpha \nu}~^{(n-1)}R_{\beta \mu}
-h_{\beta \mu}~R^{(n-1)}_{\alpha \nu}+h_{\beta \nu}~^{(n-1)}R_{\alpha \mu}\right) 
\nonumber
\\
&-\frac{1}{(n-2)(n-3)}~^{(n-1)}R\left(h_{\alpha \mu}h_{\beta \nu}
-h_{\alpha \nu}h_{\beta \mu} \right)\Big\rbrace
e^{\alpha}_{A}s^{\beta}e^{\mu}_{B}s^{\nu}
\nonumber
\\
&=\mathcal{E}_{AB}+\frac{1}{(n-3)}\left(q_{AB}~^{(n-1)}R_{\alpha \beta}s^{\alpha}s^{\beta}
+~^{(n-1)}R_{\alpha \beta}e^{\alpha}_{A}e^{\beta}_{B}\right)
\nonumber
\\
&-\frac{1}{(n-2)(n-3)}~^{(n-1)}Rq_{AB}
\end{align}
where we have defined the tensor $\mathcal{E}_{AB}=^{(n-1)}C_{\alpha \beta \mu \nu}e^{\alpha}_{A}s^{\beta}e^{\mu}_{B}s^{\nu}$. Then the effective equation on the $(n-2)$ dimensional hypersurface turns out to be,
\begin{align}\label{GCEq19}
^{(n-2)}G_{AB}&=~^{(n-1)}G_{\alpha \beta}e^{\alpha}_{A}e^{\beta}_{B}
+~^{(n-1)}R_{\alpha \beta}s^{\alpha}s^{\beta}q_{AB}-\mathcal{E}_{AB}
\nonumber
\\
&-\frac{1}{(n-3)}\left(q_{AB}~^{(n-1)}R_{\alpha \beta}s^{\alpha}s^{\beta}
+~^{(n-1)}R_{\alpha \beta}e^{\alpha}_{A}e^{\beta}_{B} \right)
+\frac{1}{(n-2)(n-3)}q_{AB}~^{(n-1)}R
\nonumber
\\
&+\left[\mathcal{K}_{AB}\mathcal{K}-\mathcal{K}_{AC}\mathcal{K}^{C}_{B}
-\frac{1}{2}q_{AB}\left(\mathcal{K}^{2}-\mathcal{K}_{AB}\mathcal{K}^{AB}\right)\right]
\nonumber
\\
&=~^{(n-1)}G_{\alpha \beta}e^{\alpha}_{A}e^{\beta}_{B}
+\frac{(n-4)}{(n-3)}q_{AB}~^{(n-1)}R_{\alpha \beta}s^{\alpha}s^{\beta}
-\frac{1}{(n-3)}~^{(n-1)}R_{\alpha \beta}e^{\alpha}_{A}e^{\beta}_{B}
\nonumber
\\
&+\frac{1}{(n-2)(n-3)}q_{AB}~^{(n-1)}R-\mathcal{E}_{AB}
\nonumber
\\
&+\left[\mathcal{K}_{AB}\mathcal{K}-\mathcal{K}_{AC}\mathcal{K}^{C}_{B}
-\frac{1}{2}q_{AB}\left(\mathcal{K}^{2}-\mathcal{K}_{CD}\mathcal{K}^{CD}\right)\right]
\end{align}
Now instead of $^{(n-1)}R_{\alpha \beta}$, we can use the following relation: $^{(n-1)}R_{\alpha \beta}=~^{(n-1)}G_{\alpha \beta}+\frac{1}{2}h_{\alpha \beta}~^{(n-1)}R$, to replace Ricci tensors in terms of Ricci scalar and Einstein tensors. This leads to,
\begin{align}\label{GCEq20}
^{(n-2)}G_{AB}&=\frac{(n-4)}{(n-3)}\left[~^{(n-1)}G_{\alpha \beta}e^{\alpha}_{A}e^{\beta}_{B}
+\left\lbrace ~^{(n-1)}G_{\alpha \beta}s^{\alpha}s^{\beta}
+\frac{(n-3)}{2(n-2)}~^{(n-1)}R \right\rbrace q_{AB}\right]
\nonumber
\\
&-\mathcal{E}_{AB}+\left[\mathcal{K}_{AB}\mathcal{K}-\mathcal{K}_{AC}\mathcal{K}^{C}_{B}
-\frac{1}{2}q_{AB}\left(\mathcal{K}^{2}-\mathcal{K}_{CD}\mathcal{K}^{CD}\right)\right]
\end{align}
Now introducing the Einstein equation in $(n-1)$ dimension we arrive at, 
\begin{equation}\label{GCEq21}
^{(n-1)}G_{\alpha \beta}=\kappa _{n-1}^{2}\mathcal{T}_{\alpha \beta};\qquad
^{(n-1)}R=-\frac{2}{(n-3)}\kappa _{n-1}^{2}\mathcal{T}
\end{equation}
Using this expression in \eq{GCEq20} leads to the effective field equation in $(n-2)$-dimension, which is presented in \sect{Sec02:n-2}.


\end{document}